\documentclass[%
reprint,
 amsmath,amssymb,
 aps,prl,
floatfix,
]{revtex4-2}

\usepackage{graphicx}% Include figure files
\usepackage{dcolumn}% Align table columns on decimal point
\usepackage{bm}% bold math
\usepackage{braket}
\usepackage{hyperref}% add hypertext capabilities
\hypersetup{pdfpagemode=FullScreen,
colorlinks=true, citecolor = blue, linkcolor=blue, filecolor=blue}
\usepackage[mathlines]{lineno}% Enable numbering of text and display math
\usepackage{amsmath}
\usepackage[T1]{fontenc}

\begin{document}

\preprint{APS/123-QED}

\title{Bench-top Cooling of a Microwave Mode using an Optically Pumped Spin Refrigerator}% Force line breaks with \\

\author{Hao Wu}

\author{Shamil Mirkhanov}%
\author{Wern Ng}
\author{Mark Oxborrow}
 \email{m.oxborrow@imperial.ac.uk}
\affiliation{%
 Department of Materials, Imperial College London, Exhibition Road, London SW7 2AZ,
 United Kingdom %\textbackslash\textbackslash
}%


%\affiliation{
%\affiliation{
 %Third institution, the second for Charlie Author
%\affiliation{%


\date{\today}% It is always \today, today,

\begin{abstract}
We experimentally demonstrate the temporary removal of  thermal photons from a microwave mode at 1.45~GHz through its interaction with the spin-polarized triplet states of photo-excited pentacene molecules doped within a \textit{p}-terphenyl crystal at room temperature. The crystal functions electromagnetically as a narrow-band cryogenic load, removing photons from the otherwise room-temperature mode via stimulated absorption. The noise temperature of the microwave mode dropped to $50^{+18}_{-32}$~K (as directly inferred by noise-power measurements) while the metal walls of the cavity enclosing the mode remained at room temperature. Simulations based on the same system's behavior as a maser (which could be characterized more accurately) indicate the possibility of the mode's temperature sinking to $\sim$10~K (corresponding to $\sim$140 microwave photons).%, below the noise floor of our room-temperature instrumentation. 
These observations, when combined with engineering improvements to deepen the cooling, identify the system as a narrow-band yet extremely convenient platform ---free of cryogenics, vacuum chambers and strong magnets--- for realizing low-noise detectors, quantum memory and quantum-enhanced machines (such as heat engines) based on strong spin-photon coupling and entanglement at microwave frequencies. 

%\begin{description}
%\end{description}
\end{abstract}

\maketitle

%\tableofcontents

%\section{\label{sec:level1}First-level %heading:\protect\\ The line
%break was forced \lowercase{via} %\textbackslash\textbackslash}

On a warm planet, electromagnetic noise associated with thermal (black--body) radiation is the ubiquitous bugbear of %precision
quantum measurements\cite{shih2018introduction} --especially %those operating at radio or 
at microwave frequencies, where a quantum of energy pales in comparison with $k_\textrm{B} T$. In a single electromagnetic mode of frequency $f_\textrm{mode}$, the mean occupation number of thermal photons is
$\bar{n}=[\exp(h f_\text{mode} /k_\textrm{B}T)-1]^{-1}$, equating to $\sim$6,200 for a microwave mode at 1~GHz at room temperature.  These photons show up as noise on any signal extracted from the mode via a coupler and %are what, for example,
can limit the speed at which an EPR/NMR spectrum or MRI scan %can be 
is taken. It is thus extremely challenging to attain the single-photon limit at microwave (or lower) frequencies with sources\cite{sieg1964microwave,benmessai2008measurement,gordon1955maser}, sensors\cite{meystre2013short,cox2018quantum} and/or detectors\cite{inomata2016single,wrachtrup2016single} 
as is necessary to implement quantum Hanbury-Brown \& Twiss %(i.e.~Hong-Ou-Mandel)
interferometry\cite{brown1956correlation,knill2001} or other more complex protocols exploiting entanglement. 

The most familiar way of removing thermal photons and their associated noise is to cool the microwave circuitry down to %a (deeply) 
cryogenic temperatures by housing it inside a dilution %(or adiabatic-demagnetization)
refrigerator.
Such refrigerators are  however bulky, mechanically fragile, and energy-guzzling (dissipating typically kilowatts during operation); these attributes alas
exclude many applications. The alternative pursued here is similar to the use, in radiometry, of a cryogenic load connected (to the rest of the microwave circuitry) through a low-loss waveguide\cite{gervasi1995}. Albeit only possible over a narrow band of frequencies, %(a limitation which admittedly also excludes applications)
the cryogenic load is here replaced by a room-temperature yet ``spin-cold'' quantum system capable of absorbing photons through stimulated absorption\cite{sieg1964microwave}. Mode-cooling is only achieved within the linewidth of the exploited quantum transition, but this cooling still enables useful applications. Our method, which exploits a cavity of  high magnetic Purcell factor\cite{breeze2015enhanced}, is thus similar in spirit to recent, cryogenic implementations of radiative cooling \cite{albanese2020radiative,xu2020}, but where the cold reservoir is a spin-bath. 

Several decades ago, it was demonstrated how Rydberg atoms could remove thermal photons from millimeter-wave cavities (operating at $\sim$100 GHz or higher) via stimulated absorption \cite{gallagher1979interactions,figger1980photon,beiting1979effects,koch1980direct,spencer1982temperature,raimond1982collective,haroche1985radiative}. The vacuum %(+ occasionally cryogenic)
equipment needed to do this was still quite bulky, however. Compared to our work at 1.45 GHz presented here, the cavity modes cooled enjoyed a
``head-start'' by containing far fewer thermal photons ($\bar{n}<50$ at room temperature\cite{raimond1982collective}) in the first place.

\begin{figure}[htbp!]
\includegraphics{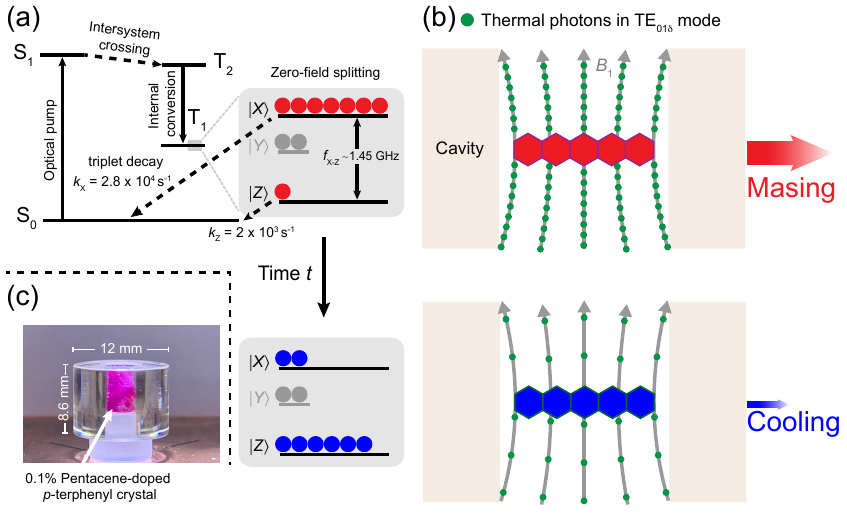}% Here is how to import EPS art
\caption{\label{fig:1} 
(a) Simplified Jablonski diagram for molecular pentacene together with how the polarization across the $\ket{\text{X}}$ and
and $\ket{\text{Z}}$ sub-levels of the lowest triplet state evolves from emissive (red circles) to absorptive (blue circles).    
(b) Dependence of the number of thermal photons (green dots) occupying the cavity's TE$_{01\delta}$ mode (grey lines of a.c.~magnetic flux) on the population distribution across the X-Z transition. 
(c) Photograph of the strontiun-titanate ring holding a 0.1\% pentacene-doped \textit{p}-terphenyl crystal. Its surrounding cylindrical copper enclosure is not shown.}
\end{figure}

%The Basic Idea
In this letter, we demonstrate a compact refrigerator capable of operating at room temperature and pressure in zero applied magnetic field (``ZF''). It exploits pentacene molecules doped within a crystal of \textit{p}-terphenyl that are spin-polarized through photo-excitation. %Though other materials (like N-V$^{-}$~diamond\cite{poklonski2007}) and other polarization methods (like radical-triplet pair mechanism\cite{blank01}) could be similarly exploited and quite possibly afford their own advantages, the high solubility (thus spatial density) of pentacene in \textit{p}-terphenyl, and (in consequence) the high spin polarization density obtainable with it, motivates its selection.
The crystal with a doping concentration of 0.1\% is placed in the bore of a dielectric ring made of crystalline strontium titanate [``STO'', see Fig.~\ref{fig:1}(c)] housed within a %concentric
cylindrical copper enclosure. The ring + enclosure support a compact TE$_{01\delta}$ microwave cavity mode whose lines
of a.c.~magnetic flux density penetrate the crystal [see Fig.~\ref{fig:1}(b)], enabling effective photon $\leftrightarrow$ spin coupling between the microwave mode and the pentacene molecules.   

%Pentacene's spin dynamics
The lowest photo-excited triplet state of pentacene in \textit{p}-terphenyl has been extensively investigated for
its use in dynamic nuclear polarization (triplet-DNP)\cite{iinuma2000high},
room-temperature masers\cite{oxborrow2012room,breeze2015enhanced,salvadori2017nanosecond,wu2020invasive,wu2020room}, quantum memory \cite{breeze2017room} and photovoltaics\cite{lubert2018identifying}.
These previous studies have focused on pentacene's spin-polarization immediately after photo-excitation. We instead exploit its opposite polarization at later times for the express purpose of mode-cooling.
Fig.~\ref{fig:1}(a) shows how the triplet state is generated:
a pentacene molecule is first excited
%, through the absorption of a pump photon at a wavelength of $\sim$590 nm, 
from its singlet ground state (S$_0$) to its first excited singlet state (S$_1$). Then, through intersystem crossing (ISC), it transfers to its first-excited triplet state (T$_2)$\cite{bogatko2016molecular}, %with a yield of $\sim$62.5\%\cite{takeda2002zero}
and thereupon undergoes rapid internal conversion down to the lowest triplet state (T$_1$) while preserving its ISC-acquired spin polarization.
In zero applied magnetic field, T$_1$ comprises three separated sub-levels: $\ket{\text{X}}$, $\ket{\text{Y}}$ and $\ket{\text{Z}}$, with initial populations %follow the ratios
0.76$\,$:$\,$0.16$\,$:$\,$0.08\cite{sloop1981electron}.
Initially, due to sizable population inversion between the $\ket{\text{X}}$ and $\ket{\text{Z}}$ sub-levels, thermal photons already occupying the cavity mode are multiplied up through stimulated emission across the X-Z transition, resulting in maser oscillation [see the upper half of Fig.~\ref{fig:1}(b)].
Because the decay rate of $\ket{\text{Z}}$ ($2\times10^3$ s$^{-1}$) is considerably slower than that of $\ket{\text{X}}$ ($2.8\times10^4$ s$^{-1}$) and the spin-lattice relaxation between them is slow compared to these rates\cite{wu2019unraveling}, the $\ket{\text{Z}}$ sub-level becomes, after masing has ceased, significantly over-populated relative to $\ket{\text{X}}$; see the bottom of Fig.~\ref{fig:1}(a).
An extremely spin-cold two-level system across $\ket{\text{X}}$ $\leftrightarrow$ $\ket{\text{Z}}$ is thus temporarily formed. As shown conceptually in the lower half of Fig.~\ref{fig:1}(b), this system will act to \textit{a}ttenuate (i.e.~remove photons from) the  electromagnetic mode through stimulated \textit{a}bsorption.
Mimicking Gordon and Townes' original %(and rather catchy) 
acronym, we refer to this refrigeration process as ``m\textit{a}s\textit{a}r" cooling.  

\begin{figure}[!htbp]
\includegraphics{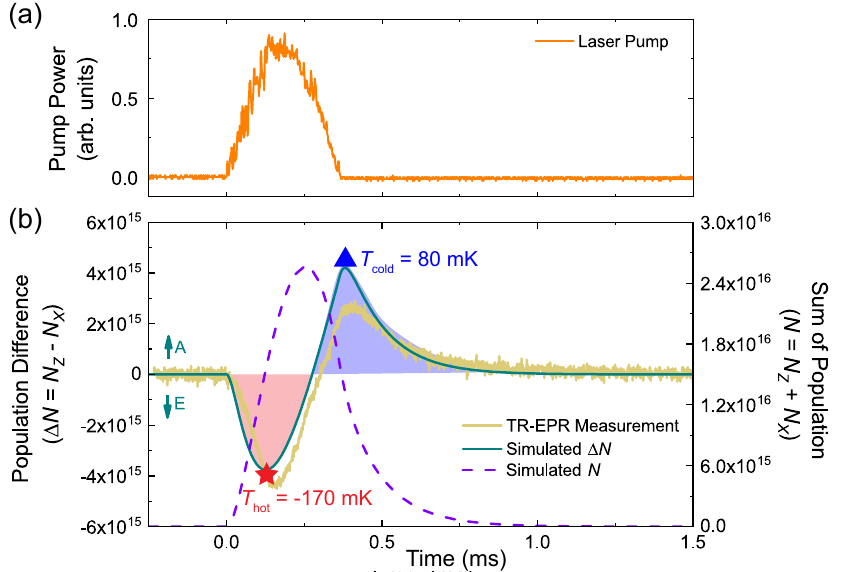}% Here is how to import EPS art
\caption{\label{fig:2} 
(a) Instantaneous pump power of dye laser as used for our TR-EPR experiment. 
(b) Measured (olive green trace) and simulated (solid teal green line) TR-EPR response of a 0.1\% pentacene-doped \textit{p}-terphenyl crystal pumped by the same dye laser. This response measures the population difference ($\Delta N$) between the triplet sub-levels, $\ket{\text{X}}$ and $\ket{\text{Z}}$. The total population ($N$) in the two sub-levels is modeled as a function of time and indicated by a dashed purple line. Red and blue shading emphasizes the emissive (E) and subsequently absorptive (A) epochs of the X-Z transition respectively.
}
\end{figure}

To verify the concept, we first quantified the achievable spin temperature, $T_\text{X-Z}$, of the system using zero-field time-resolved electron paramagnetic resonance (TR-EPR) performed at room temperature, with the pentacene-doped \textit{p}-terphenyl crystal pumped by a long-pulse dye laser. Details of this technique and how our crystal was grown have been reported elsewhere\cite{wu2019unraveling}.
Here, the spin temperature is determined by the relative instantaneous populations of the  $\ket{\text{X}}$ and $\ket{\text{Z}}$ sub-levels \cite{sieg1964microwave}:
\begin{equation}
    T_\text{X-Z} = h f_\textrm{X-Z} / [2 k_\text{B} \tanh^{-1}(\Delta N/N)],
\label{eq:1}
\end{equation}
where $f_\textrm{X-Z} = $ 1.4495 GHz is the frequency of the X-Z transition (at ZF), the population difference  $\Delta N = N_\text{Z}-N_\text{X}$, and the total population $N=N_\text{Z}+N_\text{X}$. 

Fig.~\ref{fig:2}(a) displays the time profile of a typical 590-nm pump pulse, with the dye laser set to a %relatively
low output such that no masing occurs; it lasts $\sim$300 $\mu$s and integrates to 250 mJ in energy. As displayed in Fig.~\ref{fig:2}(b), the measured signal, proportional to $\Delta N$, is strongly emissive for the first 250~$\mu$s but then becomes strongly absorptive for the next 500~$\mu$s.% or so.
This dramatic cross-over behavior, as was remarked upon almost four decades ago\cite{sloop1981electron}, is well fitted by the model reported in
ref.~\onlinecite{wu2019unraveling}. Using it, the time profile of $N$ can be accurately simulated, %and this too is 
as shown in Fig.~\ref{fig:2}(b). By substituting the simulated values of $\Delta N$ and $N$ into Eq.~(\ref{eq:1}), one calculates that the maximum excursions of the polarization in the emissive and absorptive  epochs correspond to Boltzmann-equivalent spin temperatures of -170 mK (red star) and 80 mK (blue triangle), respectively.

\begin{figure}[htbp!]
\includegraphics{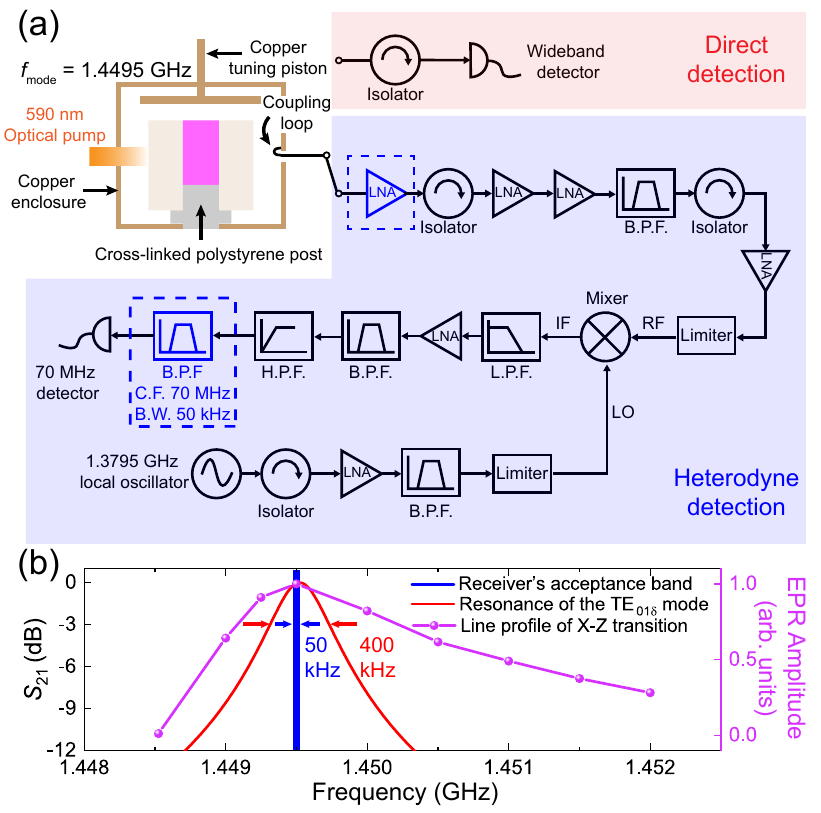}% Here is how to import EPS art
\caption{\label{fig:3} 
(a) Experimental arrangement for measuring the energy of the  TE$_{01\delta}$ mode of an STO cavity at room temperature. (Top left) STO ring (beige) housing a 0.1\% pentacene-doped \textit{p}-terphenyl crystal (pink) inside a copper enclosure (with hole for optical access). (Top right) Direct detection of the maser output with a wideband detector. (Bottom right) Heterodyne receiver suitable for measuring low mode energies. C.F.~stands for center frequency; B.P.F., H.P.F., and L.P.F.~stand for band-, high-, and low-pass filter, respectively. (b) Comparison between the receiver's acceptance band (blue) and the measured widths of the TE$_{01\delta}$ mode (red) and pentacene's X-Z transition (purple).}
\end{figure}

We have probed the extent to which the coldness of the X-Z transition can, in practice, be transferred to a target microwave mode  by measuring the instantaneous power extracted from the mode by a metal loop threaded by a small fraction of the mode's
a.c.~magnetic flux. The power so extracted from this coupling ``port'' is to first approximation proportional to the %energy of, and thus the 
number of photons occupying the mode. The port's reflection coefficient, $\Gamma_\textrm{c}^0$, was adjusted (by changing the loop size) to zero (%corresponding to 
critical coupling) in the absence of pumping. 
Our exact experimental set-up, incorporating a high-gain heterodyne receiver, is shown in  Fig.~\ref{fig:1}(c) and Fig.~\ref{fig:3}; our Supplementary Material provides additional technical details covering how %exactly the set-up's 
the noise flow is modeled and calibrated\cite{supplementarymaterial}% (taking into account the effects of impedance mismatches)
. At $f_\textrm{X-Z}$, the %target mode (viz. the 
TE$_{01\delta}$ mode of our STO-loaded cavity had a measured linewidth at critical coupling of 400 kHz, inside of which the receiver's own measurement bandwidth of 50 kHz lay (see Fig.~\ref{fig:3}b).

\begin{figure*}
\includegraphics{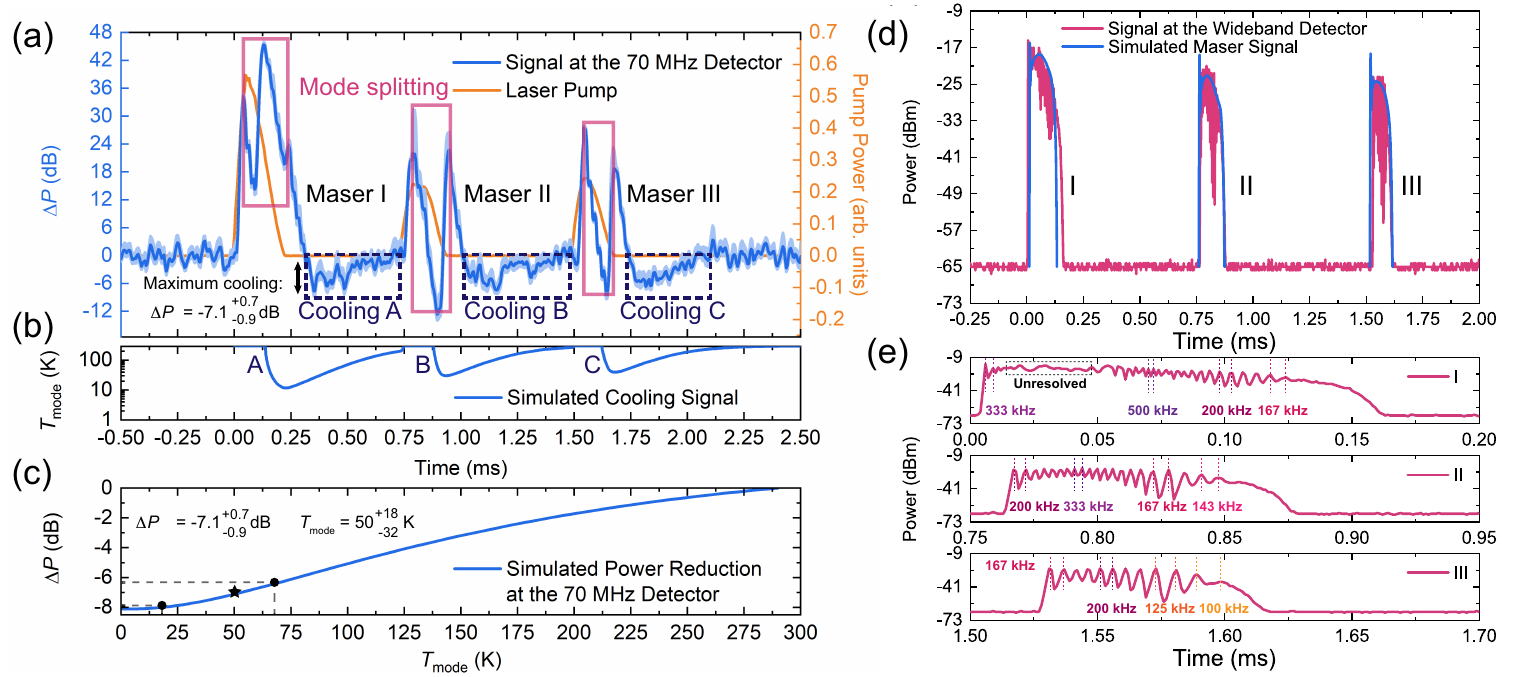}% Here is how to import EPS art
\caption{\label{fig:4} 
(a) Instantaneous power out-coupled from microwave cavity in response to a train of three optical (590 nm) pump pulses as recorded by the heterodyne receiver. The average of 11 consecutive measurements is drawn in dark blue, the standard error associated with this average is displayed as the light blue region. Masing ($\Delta P>0$) and cooling ($\Delta P<0$) signals are labeled. The average instantaneous optical pump power is shown in orange.  
(b) Simulated noise temperature of the TE$_{01\delta}$ mode as a function of time. The bright masing regime is omitted. Note that delay in the receiver's SAW filter causes the experimental traces in (a) to be delayed (by $\sim$150~$\mu$s) relative to those shown here. 
(c) Simulated power reduction ($\Delta P$) at the receiver as a function of the noise temperature of the mode $T_\textrm{mode}$. The maximum power reduction observed in (a) and its associated $T_\textrm{mode}$ (black star), 95\% upper and lower confidence limits are labeled (black circles). (d) Single trace (red) of the instantaneous out-coupled power of maser oscillation measured via the direct detection with the wideband log detector. The simulated maser signals (blue) were obtained using the same model developed for simulating the cooling signals in (b). The pump source is the same as used in (a).
(e) Zoom-in of the three maser bursts in (a) with the time-dependent Rabi frequencies labeled.
}
\end{figure*}

Our principal results are shown in Fig.~\ref{fig:4}.
The baseline at 0~dB in Fig.~\ref{fig:4}(a) includes both amplified noise received from the non-pumped cavity and the heterodyne receiver's self-generated noise. 
Excursions in the cavity's output power above this level reflect masing action; though note that, in Fig.~\ref{fig:4}(a), the peaks of the blue trace are compressed due to saturation of the later-stage amplifiers (and a protective diode limiter) in the heterodyne receiver. Dips below it arise from noise-power reduction of the microwave mode, reflecting masar cooling. 
%Here, the measured instantaneous power from the cavity is analogous to that received through a (mobile) communication channel subject to Rician fading\cite{rice58}, approaching Rayleigh fading\cite{bond1957} upon the complete demise of maser oscillation (leaving only incoherent thermal noise). The statistical characteristics of the fading (e.g., the expected number of times the instantaneous power crosses a certain level in the positive direction per second, or the expected duration of a fade in the power to a given depth) are controlled by the measurement (= channel) bandwidth, namely 50 kHz, as determined by the receiver's narrowest I.F.~filter --see Fig.~\ref{fig:3}. This bandwidth is analogous to the Doppler spread of a fading channel in mobile communications. 
Here, it is crucial to avoid mistaking the fool's gold of a ``deep fade'' (which will spontaneously occur at random times when observing narrow-band noise) from genuine cause-and-effect cooling. We suppress deep fades by averaging the recorded instantaneous power over 11 separate (statistically independent) measurements %runs
performed in quick succession; 
%this averaging implements a simple form of ``time diversity'' (in the language of communication theory); 
the resultant average, shown in dark blue in Fig.~\ref{fig:4}(a) more faithfully indicates the cooling response.

Fig.~\ref{fig:4}(a) displays that, during masing, deep notches in the out-coupled power are detected by the heterodyne receiver.
We interpret them not as cooling but instead as being caused by collective coupling between the polarized pentacene spins (regarding the X-Z transition as a two-level system) and the TE$_{01\delta}$ mode's microwave photons, causing the mode to split\cite{breeze2017room}. Temporarily, this splitting in frequency space is sufficient for the two arms of the spin-photon polariton to straddle the receiver's channel bandwidth (50 kHz), rendering them silent. Confronted by this phenomenon, the limited dynamical range (80 dB) and bandwidth of our receiver drove us to measure the out-coupled power directly with a separate wide-bandwidth log detector, shown in Fig.~\ref{fig:3}; this allowed the TE$_{01\delta}$ mode's energy to be monitored accurately during periods of strong maser bursts; traces from this detector are shown in Fig.~\ref{fig:4}(d) and (e).
The discernible Rabi oscillations in the three bursts have time-dependent frequencies ranging from $100$ to 500 kHz because the number of pentacene spins available to interact with the microwave photons varies in response to the optical pump's time profile convolved with pentacene's own spin dynamics (see Supplementary Material for details\cite{supplementarymaterial}). Oscillations faster than 500 kHz in the maser burst ``I'' are not resolved due to coarse sampling (owing to the limited memory depth of the oscilloscope used). Nevertheless, these traces show that the splitting of the TE$_{01\delta}$ mode can
certainly exceed 100~kHz, beyond the 50-kHz bandwidth of our heterodyne receiver's %steep-skirted 
band-pass %(I.F.) 
filter.

Cooling of the microwave mode is demonstrated by the received noise power dipping below its ambient (room-temperature) level after each maser burst, i.e.~the cooling epochs A to C shown in Fig.~\ref{fig:4}(a). The maximum reduction in noise power was found to be $\Delta P = -7.1^{+0.7}_{-0.9}$ dB by fitting and we here invoke a noise model based on the
``wave approach''\cite{sieg1964microwave,djordjevic2017wave}; see Supplementary Material for the error and noise analysis\cite{supplementarymaterial}.
%augmented by Mason's diagrammatic rule\cite{AgilentAN154} to quantify an etalon effect caused by reflections %of noise off the coupling loop and the front-end LNA's input (on account of their respective impedance %mismatches with the coaxial transmission line connecting the two together)%
Through this model, the relation between the mode's noise temperature (thus average photon population), $T_\textrm{mode}$, and the reduction in noise power measured at the heterodyne receiver's output, $\Delta P$, can be accurately calibrated; the curve is drawn in Fig.~\ref{fig:4}(c). As shown in Fig.~\ref{fig:4}(a), the maximum reduction in noise power observed (occurring after the first maser burst) indicates that $T_\textrm{mode}$ drops from room temperature $T_0=290$ K to $50^{+18}_{-32}$~K. This degree of cooling comes close to the limits of our instrumentation: due primarily to the input noise of the heterodyne receiver's first LNA %(plus some additional noise generated in the cable connecting this amplifier to the cavity's coupling loop)
, a dip in the noise power corresponding to $T_\textrm{mode} \le 12.5$~K cannot be discerned above the receiver's noise floor at $-8.1$~dB.

The evolution of $T_\textrm{mode}$ can also be estimated, via modeling, from the instantaneous out-coupled power as measured by the wideband detector [top of Fig.~\ref{fig:3}(a)], whose signal faithfully reflects the number of photons $q(t)$ as a function of time $t$, in the cavity during each maser burst [see Fig.~\ref{fig:4}(d)].
We point out that, similar to the heterodyne detection of extremely weak maser bursts from Rydberg atoms\cite{moi1980heterodyne}, the measured maser signals [shown in Fig.~\ref{fig:4}(a)] from our receiver suffer
both temporal delay and spread on account of its narrowest band-pass (SAW) filter. The actual duration of the maser bursts is indicated in Fig.~\ref{fig:4}(d) and (e) where the longest burst lasts about 160 $\mu$s.
The required modeling involves solving semi-classical rate equations; see Supplementary Material\cite{supplementarymaterial}.
This approach cannot simulate the Rabi oscillations observed experimentally during epochs of masing, but can accurately
predict the power envelop of each maser burst [see Fig.~\ref{fig:4}(d)]. Using the same values of fitted parameters, the same model can be used to predict the population dynamics and hence $q(t)$ during each cooling epoch [see Fig.~\ref{fig:4}(b)].
To accomplish this, the relationships $q=[\exp(h f_\text{mode} /k_\textrm{B}T_\textrm{mode})-1]^{-1}$ and $P_\textrm{maser}=q hf_\textrm{mode}\kappa_\textrm{c}k/(1+k)$\cite{breeze2017room} are used; $P_\textrm{maser}$ is the out-coupled maser power shown in Fig.~\ref{fig:4}(d), $\kappa_c = 2\pi\times0.4$ MHz is the cavity decay rate (corresponding to the cavity decay time $\sim$ 400 ns) and $k$ is the coupling coefficient of the cavity. 

Fig.~\ref{fig:4}(b) implies the cooling effect occurs almost immediately after the cessation of maser oscillation, when $\ket{\textrm{Z}}$ becomes over-populated and the rate of stimulated absorption from $\ket{\textrm{Z}}$ to $\ket{\textrm{X}}$ exceeds the rates of cavity decay (i.e. the rate of thermal photons filling in the cavity) and spin-lattice relaxation. The duration of cooling ($\sim$625 $\mu$s) is limited by the need to satisfy $N_\textrm{Z}>N_\textrm{X}$, which is controlled by the lifetime of $\ket{\textrm{Z}}$ ($\tau_{\textrm{Z}}\sim$500 $\mu$s) relative to that of $\ket{\textrm{X}}$, the former being one order of magnitude longer\cite{wu2019unraveling}. The depth of the cooling depends on the rate of stimulated absorption, which is proportional to the population difference (equal to $N_\textrm{Z}-N_\textrm{X}$) between $\ket{\textrm{X}}$ and $\ket{\textrm{Z}}$, and follows its own time evolution; see Supplementary Material\cite{supplementarymaterial}. According to the simulation, the lowest $T_\textrm{mode}$ reached was around 10~K, corresponding to $\sim140$ photons in the microwave mode. This is close to the 95\% lower confidence limit of $T_\textrm{mode}$ (i.e. $50^{+18}_{-32}$~K) inferred from our calibration curve relating $\Delta P$ to $T_\textrm{mode}$. %The discrepancy could arise from an unidentified source of additional noise.

In conclusion, our work demonstrates that
the photo-excited triplet state of pentacene doped in \textit{p}-terphenyl can be exploited
to realize a spin refrigerator that cools, by the stimulated absorption of thermal photons, an electromagnetic mode of a microwave cavity from room temperature down to a few tens of kelvin (if not lower). The cooling performance can certainly be improved upon through engineering,
e.g.~increasing the crystal's magnetic filling factor, and through material science,
i.e.~identifying (then growing) crystals exhibiting greater absorptive spin polarization.  
The approach reported here opens up a new bench-top, room-temperature route to investigating quantum entanglement\cite{haroche2006exploring} and to the realization of quantum heat engines\cite{scovil1959three,klatzow2019experimental}. A cold cavity mode so-prepared could be exploited to boost measurement sensitivity in (pulsed) EPR/NMR experiments\cite{mollier73}, radiatively cool a secondary system\cite{albanese2020radiative,xu2020}, or to reduce errors in quantum gate operations\cite{henschel2010cavity}.

\begin{acknowledgments}
We thank Ben Gaskell of Gaskell Quartz Ltd (London) for making the strontium titanate ring used. This worked was supported by the U.K. Engineering and Physical Sciences Research Council through grants  EP/K037390/1 and EP/M020398/1. H.W. acknowledges financial support from China Scholarship Council (CSC) and Imperial College London for a CSC-Imperial PhD scholarship.
\end{acknowledgments}

%\nocite{*}

%\bibliography{draftbib}% Produces the bibliography via BibTeX.

\begin{thebibliography}{50}%
\makeatletter
\providecommand \@ifxundefined [1]{%
 \@ifx{#1\undefined}
}%
\providecommand \@ifnum [1]{%
 \ifnum #1\expandafter \@firstoftwo
 \else \expandafter \@secondoftwo
 \fi
}%
\providecommand \@ifx [1]{%
 \ifx #1\expandafter \@firstoftwo
 \else \expandafter \@secondoftwo
 \fi
}%
\providecommand \natexlab [1]{#1}%
\providecommand \enquote  [1]{``#1''}%
\providecommand \bibnamefont  [1]{#1}%
\providecommand \bibfnamefont [1]{#1}%
\providecommand \citenamefont [1]{#1}%
\providecommand \href@noop [0]{\@secondoftwo}%
\providecommand \href [0]{\begingroup \@sanitize@url \@href}%
\providecommand \@href[1]{\@@startlink{#1}\@@href}%
\providecommand \@@href[1]{\endgroup#1\@@endlink}%
\providecommand \@sanitize@url [0]{\catcode `\\12\catcode `\$12\catcode
  `\&12\catcode `\#12\catcode `\^12\catcode `\_12\catcode `\%12\relax}%
\providecommand \@@startlink[1]{}%
\providecommand \@@endlink[0]{}%
\providecommand \url  [0]{\begingroup\@sanitize@url \@url }%
\providecommand \@url [1]{\endgroup\@href {#1}{\urlprefix }}%
\providecommand \urlprefix  [0]{URL }%
\providecommand \Eprint [0]{\href }%
\providecommand \doibase [0]{https://doi.org/}%
\providecommand \selectlanguage [0]{\@gobble}%
\providecommand \bibinfo  [0]{\@secondoftwo}%
\providecommand \bibfield  [0]{\@secondoftwo}%
\providecommand \translation [1]{[#1]}%
\providecommand \BibitemOpen [0]{}%
\providecommand \bibitemStop [0]{}%
\providecommand \bibitemNoStop [0]{.\EOS\space}%
\providecommand \EOS [0]{\spacefactor3000\relax}%
\providecommand \BibitemShut  [1]{\csname bibitem#1\endcsname}%
\let\auto@bib@innerbib\@empty
%</preamble>
\bibitem [{\citenamefont {Shih}(2011)}]{shih2018introduction}%
  \BibitemOpen
  \bibfield  {author} {\bibinfo {author} {\bibfnamefont {Y.}~\bibnamefont
  {Shih}},\ }\href@noop {} {\emph {\bibinfo {title} {An Introduction to Quantum
  Optics: Photon and Biphoton Physics}}}\ (\bibinfo  {publisher} {CRC Press},\
  \bibinfo {year} {2011})\BibitemShut {NoStop}%
\bibitem [{\citenamefont {Siegman}(1964)}]{sieg1964microwave}%
  \BibitemOpen
  \bibfield  {author} {\bibinfo {author} {\bibfnamefont {A.~E.}\ \bibnamefont
  {Siegman}},\ }\href@noop {} {\emph {\bibinfo {title} {Microwave Solid-State
  Masers}}}\ (\bibinfo  {publisher} {McGraw-Hill},\ \bibinfo {year}
  {1964})\BibitemShut {NoStop}%
\bibitem [{\citenamefont {Benmessai}\ \emph {et~al.}(2008)\citenamefont
  {Benmessai}, \citenamefont {Creedon}, \citenamefont {Tobar}, \citenamefont
  {Bourgeois}, \citenamefont {Kersal{\'e}},\ and\ \citenamefont
  {Giordano}}]{benmessai2008measurement}%
  \BibitemOpen
  \bibfield  {author} {\bibinfo {author} {\bibfnamefont {K.}~\bibnamefont
  {Benmessai}}, \bibinfo {author} {\bibfnamefont {D.~L.}\ \bibnamefont
  {Creedon}}, \bibinfo {author} {\bibfnamefont {M.~E.}\ \bibnamefont {Tobar}},
  \bibinfo {author} {\bibfnamefont {P.-Y.}\ \bibnamefont {Bourgeois}}, \bibinfo
  {author} {\bibfnamefont {Y.}~\bibnamefont {Kersal{\'e}}},\ and\ \bibinfo
  {author} {\bibfnamefont {V.}~\bibnamefont {Giordano}},\ }\href@noop {}
  {\bibfield  {journal} {\bibinfo  {journal} {Phys. Rev. Lett.}\ }\textbf
  {\bibinfo {volume} {100}},\ \bibinfo {pages} {233901} (\bibinfo {year}
  {2008})}\BibitemShut {NoStop}%
\bibitem [{\citenamefont {Gordon}\ \emph {et~al.}(1955)\citenamefont {Gordon},
  \citenamefont {Zeiger},\ and\ \citenamefont {Townes}}]{gordon1955maser}%
  \BibitemOpen
  \bibfield  {author} {\bibinfo {author} {\bibfnamefont {J.~P.}\ \bibnamefont
  {Gordon}}, \bibinfo {author} {\bibfnamefont {H.~J.}\ \bibnamefont {Zeiger}},\
  and\ \bibinfo {author} {\bibfnamefont {C.~H.}\ \bibnamefont {Townes}},\
  }\href@noop {} {\bibfield  {journal} {\bibinfo  {journal} {Phys. Rev.}\
  }\textbf {\bibinfo {volume} {99}},\ \bibinfo {pages} {1264} (\bibinfo {year}
  {1955})}\BibitemShut {NoStop}%
\bibitem [{\citenamefont {Meystre}(2013)}]{meystre2013short}%
  \BibitemOpen
  \bibfield  {author} {\bibinfo {author} {\bibfnamefont {P.}~\bibnamefont
  {Meystre}},\ }\href@noop {} {\bibfield  {journal} {\bibinfo  {journal} {Ann.
  Phys.}\ }\textbf {\bibinfo {volume} {525}},\ \bibinfo {pages} {215} (\bibinfo
  {year} {2013})}\BibitemShut {NoStop}%
\bibitem [{\citenamefont {Cox}\ \emph {et~al.}(2018)\citenamefont {Cox},
  \citenamefont {Meyer}, \citenamefont {Fatemi},\ and\ \citenamefont
  {Kunz}}]{cox2018quantum}%
  \BibitemOpen
  \bibfield  {author} {\bibinfo {author} {\bibfnamefont {K.~C.}\ \bibnamefont
  {Cox}}, \bibinfo {author} {\bibfnamefont {D.~H.}\ \bibnamefont {Meyer}},
  \bibinfo {author} {\bibfnamefont {F.~K.}\ \bibnamefont {Fatemi}},\ and\
  \bibinfo {author} {\bibfnamefont {P.~D.}\ \bibnamefont {Kunz}},\ }\href@noop
  {} {\bibfield  {journal} {\bibinfo  {journal} {Phys. Rev. Lett.}\ }\textbf
  {\bibinfo {volume} {121}},\ \bibinfo {pages} {110502} (\bibinfo {year}
  {2018})}\BibitemShut {NoStop}%
\bibitem [{\citenamefont {Inomata}\ \emph {et~al.}(2016)\citenamefont
  {Inomata}, \citenamefont {Lin}, \citenamefont {Koshino}, \citenamefont
  {Oliver}, \citenamefont {Tsai}, \citenamefont {Yamamoto},\ and\ \citenamefont
  {Nakamura}}]{inomata2016single}%
  \BibitemOpen
  \bibfield  {author} {\bibinfo {author} {\bibfnamefont {K.}~\bibnamefont
  {Inomata}}, \bibinfo {author} {\bibfnamefont {Z.}~\bibnamefont {Lin}},
  \bibinfo {author} {\bibfnamefont {K.}~\bibnamefont {Koshino}}, \bibinfo
  {author} {\bibfnamefont {W.~D.}\ \bibnamefont {Oliver}}, \bibinfo {author}
  {\bibfnamefont {J.-S.}\ \bibnamefont {Tsai}}, \bibinfo {author}
  {\bibfnamefont {T.}~\bibnamefont {Yamamoto}},\ and\ \bibinfo {author}
  {\bibfnamefont {Y.}~\bibnamefont {Nakamura}},\ }\href@noop {} {\bibfield
  {journal} {\bibinfo  {journal} {Nat. Commun.}\ }\textbf {\bibinfo {volume}
  {7}},\ \bibinfo {pages} {12303} (\bibinfo {year} {2016})}\BibitemShut
  {NoStop}%
\bibitem [{\citenamefont {Wrachtrup}\ and\ \citenamefont
  {Finkler}(2016)}]{wrachtrup2016single}%
  \BibitemOpen
  \bibfield  {author} {\bibinfo {author} {\bibfnamefont {J.}~\bibnamefont
  {Wrachtrup}}\ and\ \bibinfo {author} {\bibfnamefont {A.}~\bibnamefont
  {Finkler}},\ }\href@noop {} {\bibfield  {journal} {\bibinfo  {journal} {J.
  Magn. Reson.}\ }\textbf {\bibinfo {volume} {269}},\ \bibinfo {pages} {225}
  (\bibinfo {year} {2016})}\BibitemShut {NoStop}%
\bibitem [{\citenamefont {Hanbury-Brown}\ and\ \citenamefont
  {Twiss}(1956)}]{brown1956correlation}%
  \BibitemOpen
  \bibfield  {author} {\bibinfo {author} {\bibfnamefont {R.}~\bibnamefont
  {Hanbury-Brown}}\ and\ \bibinfo {author} {\bibfnamefont {R.~Q.}\ \bibnamefont
  {Twiss}},\ }\href@noop {} {\bibfield  {journal} {\bibinfo  {journal} {Nature
  (London)}\ }\textbf {\bibinfo {volume} {177}},\ \bibinfo {pages} {27}
  (\bibinfo {year} {1956})}\BibitemShut {NoStop}%
\bibitem [{\citenamefont {Knill}\ \emph {et~al.}(2001)\citenamefont {Knill},
  \citenamefont {Laflamme},\ and\ \citenamefont {Milburn}}]{knill2001}%
  \BibitemOpen
  \bibfield  {author} {\bibinfo {author} {\bibfnamefont {E.}~\bibnamefont
  {Knill}}, \bibinfo {author} {\bibfnamefont {R.}~\bibnamefont {Laflamme}},\
  and\ \bibinfo {author} {\bibfnamefont {G.}~\bibnamefont {Milburn}},\
  }\href@noop {} {\bibfield  {journal} {\bibinfo  {journal} {Nature}\ }\textbf
  {\bibinfo {volume} {409}},\ \bibinfo {pages} {46} (\bibinfo {year}
  {2001})}\BibitemShut {NoStop}%
\bibitem [{\citenamefont {Gervasi}\ \emph {et~al.}(1995)\citenamefont
  {Gervasi}, \citenamefont {Bonelli}, \citenamefont {Sironi}, \citenamefont
  {F.}, \citenamefont {Passerini},\ and\ \citenamefont {Casani}}]{gervasi1995}%
  \BibitemOpen
  \bibfield  {author} {\bibinfo {author} {\bibfnamefont {M.}~\bibnamefont
  {Gervasi}}, \bibinfo {author} {\bibfnamefont {G.}~\bibnamefont {Bonelli}},
  \bibinfo {author} {\bibfnamefont {G.}~\bibnamefont {Sironi}}, \bibinfo
  {author} {\bibfnamefont {C.}~\bibnamefont {F.}}, \bibinfo {author}
  {\bibfnamefont {A.}~\bibnamefont {Passerini}},\ and\ \bibinfo {author}
  {\bibfnamefont {S.}~\bibnamefont {Casani}},\ }\href@noop {} {\bibfield
  {journal} {\bibinfo  {journal} {Rev. Sci. Instr.}\ }\textbf {\bibinfo
  {volume} {66}},\ \bibinfo {pages} {4798} (\bibinfo {year}
  {1995})}\BibitemShut {NoStop}%
\bibitem [{\citenamefont {Breeze}\ \emph {et~al.}(2015)\citenamefont {Breeze},
  \citenamefont {Tan}, \citenamefont {Richards}, \citenamefont {Sathian},
  \citenamefont {Oxborrow},\ and\ \citenamefont {Alford}}]{breeze2015enhanced}%
  \BibitemOpen
  \bibfield  {author} {\bibinfo {author} {\bibfnamefont {J.~D.}\ \bibnamefont
  {Breeze}}, \bibinfo {author} {\bibfnamefont {K.-J.}\ \bibnamefont {Tan}},
  \bibinfo {author} {\bibfnamefont {B.}~\bibnamefont {Richards}}, \bibinfo
  {author} {\bibfnamefont {J.}~\bibnamefont {Sathian}}, \bibinfo {author}
  {\bibfnamefont {M.}~\bibnamefont {Oxborrow}},\ and\ \bibinfo {author}
  {\bibfnamefont {N.~M.}\ \bibnamefont {Alford}},\ }\href@noop {} {\bibfield
  {journal} {\bibinfo  {journal} {Nat. Commun.}\ }\textbf {\bibinfo {volume}
  {6}},\ \bibinfo {pages} {6215} (\bibinfo {year} {2015})}\BibitemShut
  {NoStop}%
\bibitem [{\citenamefont {Albanese}\ \emph {et~al.}(2020)\citenamefont
  {Albanese}, \citenamefont {Probst}, \citenamefont {Ranjan}, \citenamefont
  {Zollitsch}, \citenamefont {Pechal}, \citenamefont {Wallraff}, \citenamefont
  {Morton}, \citenamefont {Vion}, \citenamefont {Esteve}, \citenamefont
  {Flurin} \emph {et~al.}}]{albanese2020radiative}%
  \BibitemOpen
  \bibfield  {author} {\bibinfo {author} {\bibfnamefont {B.}~\bibnamefont
  {Albanese}}, \bibinfo {author} {\bibfnamefont {S.}~\bibnamefont {Probst}},
  \bibinfo {author} {\bibfnamefont {V.}~\bibnamefont {Ranjan}}, \bibinfo
  {author} {\bibfnamefont {C.~W.}\ \bibnamefont {Zollitsch}}, \bibinfo {author}
  {\bibfnamefont {M.}~\bibnamefont {Pechal}}, \bibinfo {author} {\bibfnamefont
  {A.}~\bibnamefont {Wallraff}}, \bibinfo {author} {\bibfnamefont {J.~J.}\
  \bibnamefont {Morton}}, \bibinfo {author} {\bibfnamefont {D.}~\bibnamefont
  {Vion}}, \bibinfo {author} {\bibfnamefont {D.}~\bibnamefont {Esteve}},
  \bibinfo {author} {\bibfnamefont {E.}~\bibnamefont {Flurin}}, \emph
  {et~al.},\ }\href@noop {} {\bibfield  {journal} {\bibinfo  {journal} {Nat.
  Phys.}\ }\textbf {\bibinfo {volume} {16}},\ \bibinfo {pages} {751} (\bibinfo
  {year} {2020})}\BibitemShut {NoStop}%
\bibitem [{\citenamefont {Xu}\ \emph {et~al.}(2020)\citenamefont {Xu},
  \citenamefont {Han}, \citenamefont {Zou}, \citenamefont {Fu}, \citenamefont
  {Xu}, \citenamefont {Zhong}, \citenamefont {Jiang},\ and\ \citenamefont
  {Tang}}]{xu2020}%
  \BibitemOpen
  \bibfield  {author} {\bibinfo {author} {\bibfnamefont {M.}~\bibnamefont
  {Xu}}, \bibinfo {author} {\bibfnamefont {X.}~\bibnamefont {Han}}, \bibinfo
  {author} {\bibfnamefont {C.-L.}\ \bibnamefont {Zou}}, \bibinfo {author}
  {\bibfnamefont {W.}~\bibnamefont {Fu}}, \bibinfo {author} {\bibfnamefont
  {Y.}~\bibnamefont {Xu}}, \bibinfo {author} {\bibfnamefont {C.}~\bibnamefont
  {Zhong}}, \bibinfo {author} {\bibfnamefont {L.}~\bibnamefont {Jiang}},\ and\
  \bibinfo {author} {\bibfnamefont {H.~X.}\ \bibnamefont {Tang}},\ }\href@noop
  {} {\bibfield  {journal} {\bibinfo  {journal} {Phys. Rev. Lett.}\ }\textbf
  {\bibinfo {volume} {124}},\ \bibinfo {pages} {033602} (\bibinfo {year}
  {2020})}\BibitemShut {NoStop}%
\bibitem [{\citenamefont {Gallagher}\ and\ \citenamefont
  {Cooke}(1979)}]{gallagher1979interactions}%
  \BibitemOpen
  \bibfield  {author} {\bibinfo {author} {\bibfnamefont {T.}~\bibnamefont
  {Gallagher}}\ and\ \bibinfo {author} {\bibfnamefont {W.}~\bibnamefont
  {Cooke}},\ }\href@noop {} {\bibfield  {journal} {\bibinfo  {journal} {Phys.
  Rev. Lett.}\ }\textbf {\bibinfo {volume} {42}},\ \bibinfo {pages} {835}
  (\bibinfo {year} {1979})}\BibitemShut {NoStop}%
\bibitem [{\citenamefont {Figger}\ \emph {et~al.}(1980)\citenamefont {Figger},
  \citenamefont {Leuchs}, \citenamefont {Straubinger},\ and\ \citenamefont
  {Walther}}]{figger1980photon}%
  \BibitemOpen
  \bibfield  {author} {\bibinfo {author} {\bibfnamefont {H.}~\bibnamefont
  {Figger}}, \bibinfo {author} {\bibfnamefont {G.}~\bibnamefont {Leuchs}},
  \bibinfo {author} {\bibfnamefont {R.}~\bibnamefont {Straubinger}},\ and\
  \bibinfo {author} {\bibfnamefont {H.}~\bibnamefont {Walther}},\ }\href@noop
  {} {\bibfield  {journal} {\bibinfo  {journal} {Opt. Commun.}\ }\textbf
  {\bibinfo {volume} {33}},\ \bibinfo {pages} {37} (\bibinfo {year}
  {1980})}\BibitemShut {NoStop}%
\bibitem [{\citenamefont {Beiting}\ \emph {et~al.}(1979)\citenamefont
  {Beiting}, \citenamefont {Hildebrandt}, \citenamefont {Kellert},
  \citenamefont {Foltz}, \citenamefont {Smith}, \citenamefont {Dunning},\ and\
  \citenamefont {Stebbings}}]{beiting1979effects}%
  \BibitemOpen
  \bibfield  {author} {\bibinfo {author} {\bibfnamefont {E.}~\bibnamefont
  {Beiting}}, \bibinfo {author} {\bibfnamefont {G.}~\bibnamefont
  {Hildebrandt}}, \bibinfo {author} {\bibfnamefont {F.}~\bibnamefont
  {Kellert}}, \bibinfo {author} {\bibfnamefont {G.}~\bibnamefont {Foltz}},
  \bibinfo {author} {\bibfnamefont {K.}~\bibnamefont {Smith}}, \bibinfo
  {author} {\bibfnamefont {F.}~\bibnamefont {Dunning}},\ and\ \bibinfo {author}
  {\bibfnamefont {R.}~\bibnamefont {Stebbings}},\ }\href@noop {} {\bibfield
  {journal} {\bibinfo  {journal} {J. Chem. Phys.}\ }\textbf {\bibinfo {volume}
  {70}},\ \bibinfo {pages} {3551} (\bibinfo {year} {1979})}\BibitemShut
  {NoStop}%
\bibitem [{\citenamefont {Koch}\ \emph {et~al.}(1980)\citenamefont {Koch},
  \citenamefont {Hieronymus}, \citenamefont {Van~Raan},\ and\ \citenamefont
  {Raith}}]{koch1980direct}%
  \BibitemOpen
  \bibfield  {author} {\bibinfo {author} {\bibfnamefont {P.}~\bibnamefont
  {Koch}}, \bibinfo {author} {\bibfnamefont {H.}~\bibnamefont {Hieronymus}},
  \bibinfo {author} {\bibfnamefont {A.}~\bibnamefont {Van~Raan}},\ and\
  \bibinfo {author} {\bibfnamefont {W.}~\bibnamefont {Raith}},\ }\href@noop {}
  {\bibfield  {journal} {\bibinfo  {journal} {Phys. Lett. A}\ }\textbf
  {\bibinfo {volume} {75}},\ \bibinfo {pages} {273} (\bibinfo {year}
  {1980})}\BibitemShut {NoStop}%
\bibitem [{\citenamefont {Spencer}\ \emph {et~al.}(1982)\citenamefont
  {Spencer}, \citenamefont {Vaidyanathan}, \citenamefont {Kleppner},\ and\
  \citenamefont {Ducas}}]{spencer1982temperature}%
  \BibitemOpen
  \bibfield  {author} {\bibinfo {author} {\bibfnamefont {W.~P.}\ \bibnamefont
  {Spencer}}, \bibinfo {author} {\bibfnamefont {A.~G.}\ \bibnamefont
  {Vaidyanathan}}, \bibinfo {author} {\bibfnamefont {D.}~\bibnamefont
  {Kleppner}},\ and\ \bibinfo {author} {\bibfnamefont {T.~W.}\ \bibnamefont
  {Ducas}},\ }\href@noop {} {\bibfield  {journal} {\bibinfo  {journal} {Phys.
  Rev. A}\ }\textbf {\bibinfo {volume} {25}},\ \bibinfo {pages} {380} (\bibinfo
  {year} {1982})}\BibitemShut {NoStop}%
\bibitem [{\citenamefont {Raimond}\ \emph {et~al.}(1982)\citenamefont
  {Raimond}, \citenamefont {Goy}, \citenamefont {Gross}, \citenamefont
  {Fabre},\ and\ \citenamefont {Haroche}}]{raimond1982collective}%
  \BibitemOpen
  \bibfield  {author} {\bibinfo {author} {\bibfnamefont {J.}~\bibnamefont
  {Raimond}}, \bibinfo {author} {\bibfnamefont {P.}~\bibnamefont {Goy}},
  \bibinfo {author} {\bibfnamefont {M.}~\bibnamefont {Gross}}, \bibinfo
  {author} {\bibfnamefont {C.}~\bibnamefont {Fabre}},\ and\ \bibinfo {author}
  {\bibfnamefont {S.}~\bibnamefont {Haroche}},\ }\href@noop {} {\bibfield
  {journal} {\bibinfo  {journal} {Phys. Rev. Lett.}\ }\textbf {\bibinfo
  {volume} {49}},\ \bibinfo {pages} {117} (\bibinfo {year} {1982})}\BibitemShut
  {NoStop}%
\bibitem [{\citenamefont {Haroche}\ and\ \citenamefont
  {Raimond}(1985)}]{haroche1985radiative}%
  \BibitemOpen
  \bibfield  {author} {\bibinfo {author} {\bibfnamefont {S.}~\bibnamefont
  {Haroche}}\ and\ \bibinfo {author} {\bibfnamefont {J.}~\bibnamefont
  {Raimond}},\ }\href@noop {} {\bibfield  {journal} {\bibinfo  {journal} {Adv.
  At. Mol. Phys.}\ }\textbf {\bibinfo {volume} {20}},\ \bibinfo {pages} {347}
  (\bibinfo {year} {1985})}\BibitemShut {NoStop}%
\bibitem [{\citenamefont {Iinuma}\ \emph {et~al.}(2000)\citenamefont {Iinuma},
  \citenamefont {Takahashi}, \citenamefont {Shak{\'e}}, \citenamefont {Oda},
  \citenamefont {Masaike}, \citenamefont {Yabuzaki},\ and\ \citenamefont
  {Shimizu}}]{iinuma2000high}%
  \BibitemOpen
  \bibfield  {author} {\bibinfo {author} {\bibfnamefont {M.}~\bibnamefont
  {Iinuma}}, \bibinfo {author} {\bibfnamefont {Y.}~\bibnamefont {Takahashi}},
  \bibinfo {author} {\bibfnamefont {I.}~\bibnamefont {Shak{\'e}}}, \bibinfo
  {author} {\bibfnamefont {M.}~\bibnamefont {Oda}}, \bibinfo {author}
  {\bibfnamefont {A.}~\bibnamefont {Masaike}}, \bibinfo {author} {\bibfnamefont
  {T.}~\bibnamefont {Yabuzaki}},\ and\ \bibinfo {author} {\bibfnamefont
  {H.~M.}\ \bibnamefont {Shimizu}},\ }\href@noop {} {\bibfield  {journal}
  {\bibinfo  {journal} {Phys. Rev. Lett.}\ }\textbf {\bibinfo {volume} {84}},\
  \bibinfo {pages} {171} (\bibinfo {year} {2000})}\BibitemShut {NoStop}%
\bibitem [{\citenamefont {Oxborrow}\ \emph {et~al.}(2012)\citenamefont
  {Oxborrow}, \citenamefont {Breeze},\ and\ \citenamefont
  {Alford}}]{oxborrow2012room}%
  \BibitemOpen
  \bibfield  {author} {\bibinfo {author} {\bibfnamefont {M.}~\bibnamefont
  {Oxborrow}}, \bibinfo {author} {\bibfnamefont {J.~D.}\ \bibnamefont
  {Breeze}},\ and\ \bibinfo {author} {\bibfnamefont {N.~M.}\ \bibnamefont
  {Alford}},\ }\href@noop {} {\bibfield  {journal} {\bibinfo  {journal} {Nature
  (London)}\ }\textbf {\bibinfo {volume} {488}},\ \bibinfo {pages} {353}
  (\bibinfo {year} {2012})}\BibitemShut {NoStop}%
\bibitem [{\citenamefont {Salvadori}\ \emph {et~al.}(2017)\citenamefont
  {Salvadori}, \citenamefont {Breeze}, \citenamefont {Tan}, \citenamefont
  {Sathian}, \citenamefont {Richards}, \citenamefont {Fung}, \citenamefont
  {Wolfowicz}, \citenamefont {Oxborrow}, \citenamefont {Alford},\ and\
  \citenamefont {Kay}}]{salvadori2017nanosecond}%
  \BibitemOpen
  \bibfield  {author} {\bibinfo {author} {\bibfnamefont {E.}~\bibnamefont
  {Salvadori}}, \bibinfo {author} {\bibfnamefont {J.~D.}\ \bibnamefont
  {Breeze}}, \bibinfo {author} {\bibfnamefont {K.-J.}\ \bibnamefont {Tan}},
  \bibinfo {author} {\bibfnamefont {J.}~\bibnamefont {Sathian}}, \bibinfo
  {author} {\bibfnamefont {B.}~\bibnamefont {Richards}}, \bibinfo {author}
  {\bibfnamefont {M.~W.}\ \bibnamefont {Fung}}, \bibinfo {author}
  {\bibfnamefont {G.}~\bibnamefont {Wolfowicz}}, \bibinfo {author}
  {\bibfnamefont {M.}~\bibnamefont {Oxborrow}}, \bibinfo {author}
  {\bibfnamefont {N.~M.}\ \bibnamefont {Alford}},\ and\ \bibinfo {author}
  {\bibfnamefont {C.~W.}\ \bibnamefont {Kay}},\ }\href@noop {} {\bibfield
  {journal} {\bibinfo  {journal} {Sci. Rep.}\ }\textbf {\bibinfo {volume}
  {7}},\ \bibinfo {pages} {1} (\bibinfo {year} {2017})}\BibitemShut {NoStop}%
\bibitem [{\citenamefont {Wu}\ \emph {et~al.}(2020{\natexlab{a}})\citenamefont
  {Wu}, \citenamefont {Mirkhanov}, \citenamefont {Ng}, \citenamefont {Chen},
  \citenamefont {Xiong},\ and\ \citenamefont {Oxborrow}}]{wu2020invasive}%
  \BibitemOpen
  \bibfield  {author} {\bibinfo {author} {\bibfnamefont {H.}~\bibnamefont
  {Wu}}, \bibinfo {author} {\bibfnamefont {S.}~\bibnamefont {Mirkhanov}},
  \bibinfo {author} {\bibfnamefont {W.}~\bibnamefont {Ng}}, \bibinfo {author}
  {\bibfnamefont {K.-C.}\ \bibnamefont {Chen}}, \bibinfo {author}
  {\bibfnamefont {Y.}~\bibnamefont {Xiong}},\ and\ \bibinfo {author}
  {\bibfnamefont {M.}~\bibnamefont {Oxborrow}},\ }\href@noop {} {\bibfield
  {journal} {\bibinfo  {journal} {Opt. Express}\ }\textbf {\bibinfo {volume}
  {28}},\ \bibinfo {pages} {29691} (\bibinfo {year}
  {2020}{\natexlab{a}})}\BibitemShut {NoStop}%
\bibitem [{\citenamefont {Wu}\ \emph {et~al.}(2020{\natexlab{b}})\citenamefont
  {Wu}, \citenamefont {Xie}, \citenamefont {Ng}, \citenamefont {Mehanna},
  \citenamefont {Li}, \citenamefont {Attwood},\ and\ \citenamefont
  {Oxborrow}}]{wu2020room}%
  \BibitemOpen
  \bibfield  {author} {\bibinfo {author} {\bibfnamefont {H.}~\bibnamefont
  {Wu}}, \bibinfo {author} {\bibfnamefont {X.}~\bibnamefont {Xie}}, \bibinfo
  {author} {\bibfnamefont {W.}~\bibnamefont {Ng}}, \bibinfo {author}
  {\bibfnamefont {S.}~\bibnamefont {Mehanna}}, \bibinfo {author} {\bibfnamefont
  {Y.}~\bibnamefont {Li}}, \bibinfo {author} {\bibfnamefont {M.}~\bibnamefont
  {Attwood}},\ and\ \bibinfo {author} {\bibfnamefont {M.}~\bibnamefont
  {Oxborrow}},\ }\href@noop {} {\bibfield  {journal} {\bibinfo  {journal}
  {Phys. Rev. Appl.}\ }\textbf {\bibinfo {volume} {14}},\ \bibinfo {pages}
  {064017} (\bibinfo {year} {2020}{\natexlab{b}})}\BibitemShut {NoStop}%
\bibitem [{\citenamefont {Breeze}\ \emph {et~al.}(2017)\citenamefont {Breeze},
  \citenamefont {Salvadori}, \citenamefont {Sathian}, \citenamefont {Alford},\
  and\ \citenamefont {Kay}}]{breeze2017room}%
  \BibitemOpen
  \bibfield  {author} {\bibinfo {author} {\bibfnamefont {J.~D.}\ \bibnamefont
  {Breeze}}, \bibinfo {author} {\bibfnamefont {E.}~\bibnamefont {Salvadori}},
  \bibinfo {author} {\bibfnamefont {J.}~\bibnamefont {Sathian}}, \bibinfo
  {author} {\bibfnamefont {N.~M.}\ \bibnamefont {Alford}},\ and\ \bibinfo
  {author} {\bibfnamefont {C.~W.}\ \bibnamefont {Kay}},\ }\href@noop {}
  {\bibfield  {journal} {\bibinfo  {journal} {npj Quantum Inf.}\ }\textbf
  {\bibinfo {volume} {3}},\ \bibinfo {pages} {40} (\bibinfo {year}
  {2017})}\BibitemShut {NoStop}%
\bibitem [{\citenamefont {Lubert-Perquel}\ \emph {et~al.}(2018)\citenamefont
  {Lubert-Perquel}, \citenamefont {Salvadori}, \citenamefont {Dyson},
  \citenamefont {Stavrinou}, \citenamefont {Montis}, \citenamefont {Nagashima},
  \citenamefont {Kobori}, \citenamefont {Heutz},\ and\ \citenamefont
  {Kay}}]{lubert2018identifying}%
  \BibitemOpen
  \bibfield  {author} {\bibinfo {author} {\bibfnamefont {D.}~\bibnamefont
  {Lubert-Perquel}}, \bibinfo {author} {\bibfnamefont {E.}~\bibnamefont
  {Salvadori}}, \bibinfo {author} {\bibfnamefont {M.}~\bibnamefont {Dyson}},
  \bibinfo {author} {\bibfnamefont {P.~N.}\ \bibnamefont {Stavrinou}}, \bibinfo
  {author} {\bibfnamefont {R.}~\bibnamefont {Montis}}, \bibinfo {author}
  {\bibfnamefont {H.}~\bibnamefont {Nagashima}}, \bibinfo {author}
  {\bibfnamefont {Y.}~\bibnamefont {Kobori}}, \bibinfo {author} {\bibfnamefont
  {S.}~\bibnamefont {Heutz}},\ and\ \bibinfo {author} {\bibfnamefont {C.~W.}\
  \bibnamefont {Kay}},\ }\href@noop {} {\bibfield  {journal} {\bibinfo
  {journal} {Nat. Commun.}\ }\textbf {\bibinfo {volume} {9}},\ \bibinfo {pages}
  {4222} (\bibinfo {year} {2018})}\BibitemShut {NoStop}%
\bibitem [{\citenamefont {Bogatko}\ \emph {et~al.}(2016)\citenamefont
  {Bogatko}, \citenamefont {Haynes}, \citenamefont {Sathian}, \citenamefont
  {Wade}, \citenamefont {Kim}, \citenamefont {Tan}, \citenamefont {Breeze},
  \citenamefont {Salvadori}, \citenamefont {Horsfield},\ and\ \citenamefont
  {Oxborrow}}]{bogatko2016molecular}%
  \BibitemOpen
  \bibfield  {author} {\bibinfo {author} {\bibfnamefont {S.}~\bibnamefont
  {Bogatko}}, \bibinfo {author} {\bibfnamefont {P.~D.}\ \bibnamefont {Haynes}},
  \bibinfo {author} {\bibfnamefont {J.}~\bibnamefont {Sathian}}, \bibinfo
  {author} {\bibfnamefont {J.}~\bibnamefont {Wade}}, \bibinfo {author}
  {\bibfnamefont {J.-S.}\ \bibnamefont {Kim}}, \bibinfo {author} {\bibfnamefont
  {K.-J.}\ \bibnamefont {Tan}}, \bibinfo {author} {\bibfnamefont {J.~D.}\
  \bibnamefont {Breeze}}, \bibinfo {author} {\bibfnamefont {E.}~\bibnamefont
  {Salvadori}}, \bibinfo {author} {\bibfnamefont {A.}~\bibnamefont
  {Horsfield}},\ and\ \bibinfo {author} {\bibfnamefont {M.}~\bibnamefont
  {Oxborrow}},\ }\href@noop {} {\bibfield  {journal} {\bibinfo  {journal} {J.
  Phys. Chem. C}\ }\textbf {\bibinfo {volume} {120}},\ \bibinfo {pages} {8251}
  (\bibinfo {year} {2016})}\BibitemShut {NoStop}%
\bibitem [{\citenamefont {Sloop}\ \emph {et~al.}(1981)\citenamefont {Sloop},
  \citenamefont {Yu}, \citenamefont {Lin},\ and\ \citenamefont
  {Weissman}}]{sloop1981electron}%
  \BibitemOpen
  \bibfield  {author} {\bibinfo {author} {\bibfnamefont {D.~J.}\ \bibnamefont
  {Sloop}}, \bibinfo {author} {\bibfnamefont {H.-L.}\ \bibnamefont {Yu}},
  \bibinfo {author} {\bibfnamefont {T.-S.}\ \bibnamefont {Lin}},\ and\ \bibinfo
  {author} {\bibfnamefont {S.}~\bibnamefont {Weissman}},\ }\href@noop {}
  {\bibfield  {journal} {\bibinfo  {journal} {J. Chem. Phys.}\ }\textbf
  {\bibinfo {volume} {75}},\ \bibinfo {pages} {3746} (\bibinfo {year}
  {1981})}\BibitemShut {NoStop}%
\bibitem [{\citenamefont {Wu}\ \emph {et~al.}(2019)\citenamefont {Wu},
  \citenamefont {Ng}, \citenamefont {Mirkhanov}, \citenamefont {Amirzhan},
  \citenamefont {Nitnara},\ and\ \citenamefont {Oxborrow}}]{wu2019unraveling}%
  \BibitemOpen
  \bibfield  {author} {\bibinfo {author} {\bibfnamefont {H.}~\bibnamefont
  {Wu}}, \bibinfo {author} {\bibfnamefont {W.}~\bibnamefont {Ng}}, \bibinfo
  {author} {\bibfnamefont {S.}~\bibnamefont {Mirkhanov}}, \bibinfo {author}
  {\bibfnamefont {A.}~\bibnamefont {Amirzhan}}, \bibinfo {author}
  {\bibfnamefont {S.}~\bibnamefont {Nitnara}},\ and\ \bibinfo {author}
  {\bibfnamefont {M.}~\bibnamefont {Oxborrow}},\ }\href@noop {} {\bibfield
  {journal} {\bibinfo  {journal} {J. Phys. Chem. C}\ }\textbf {\bibinfo
  {volume} {123}},\ \bibinfo {pages} {24275} (\bibinfo {year}
  {2019})}\BibitemShut {NoStop}%
\bibitem [{sup()}]{supplementarymaterial}%
  \BibitemOpen
  \bibfield  {title} {\bibinfo {title} {{See Supplementary Material at [url]
  for details of experimental method, model of pentacene's spin dynamics and
  noise analysis, which includes
  Refs}.~\cite{oxborrow2007,wu2020room,moi1980heterodyne,pollnau1994explanation,patterson1984b,takeda2002zero,sloop1981electron,wu2019unraveling,nelson1981laser,deeg1985,sieg1964microwave,salvadori2017nanosecond,yang2000zero,penfield1962noisewave,djordjevic2017wave,PMA2,kraus1986radio,oxborrow2017maser}},\
  }\href@noop {} {\ }\BibitemShut {NoStop}%
\bibitem [{\citenamefont {{\DJ}or{\dj}evi{\'c}}\ \emph
  {et~al.}(2017)\citenamefont {{\DJ}or{\dj}evi{\'c}}, \citenamefont
  {Marinkovi{\'c}}, \citenamefont {Crupi}, \citenamefont
  {Proni{\'c}-Ran{\v{c}}i{\'c}}, \citenamefont {Markovi{\'c}},\ and\
  \citenamefont {Caddemi}}]{djordjevic2017wave}%
  \BibitemOpen
  \bibfield  {author} {\bibinfo {author} {\bibfnamefont {V.}~\bibnamefont
  {{\DJ}or{\dj}evi{\'c}}}, \bibinfo {author} {\bibfnamefont {Z.}~\bibnamefont
  {Marinkovi{\'c}}}, \bibinfo {author} {\bibfnamefont {G.}~\bibnamefont
  {Crupi}}, \bibinfo {author} {\bibfnamefont {O.}~\bibnamefont
  {Proni{\'c}-Ran{\v{c}}i{\'c}}}, \bibinfo {author} {\bibfnamefont
  {V.}~\bibnamefont {Markovi{\'c}}},\ and\ \bibinfo {author} {\bibfnamefont
  {A.}~\bibnamefont {Caddemi}},\ }\href@noop {} {\bibfield  {journal} {\bibinfo
   {journal} {Int. J. Numer. Model.}\ }\textbf {\bibinfo {volume} {30}},\
  \bibinfo {pages} {e2138} (\bibinfo {year} {2017})}\BibitemShut {NoStop}%
\bibitem [{\citenamefont {Moi}\ \emph {et~al.}(1980)\citenamefont {Moi},
  \citenamefont {Fabre}, \citenamefont {Goy}, \citenamefont {Gross},
  \citenamefont {Haroche}, \citenamefont {Encrenaz}, \citenamefont {Beaudin},\
  and\ \citenamefont {Lazareff}}]{moi1980heterodyne}%
  \BibitemOpen
  \bibfield  {author} {\bibinfo {author} {\bibfnamefont {L.}~\bibnamefont
  {Moi}}, \bibinfo {author} {\bibfnamefont {C.}~\bibnamefont {Fabre}}, \bibinfo
  {author} {\bibfnamefont {P.}~\bibnamefont {Goy}}, \bibinfo {author}
  {\bibfnamefont {M.}~\bibnamefont {Gross}}, \bibinfo {author} {\bibfnamefont
  {S.}~\bibnamefont {Haroche}}, \bibinfo {author} {\bibfnamefont
  {P.}~\bibnamefont {Encrenaz}}, \bibinfo {author} {\bibfnamefont
  {G.}~\bibnamefont {Beaudin}},\ and\ \bibinfo {author} {\bibfnamefont
  {B.}~\bibnamefont {Lazareff}},\ }\href@noop {} {\bibfield  {journal}
  {\bibinfo  {journal} {Opt. Commun.}\ }\textbf {\bibinfo {volume} {33}},\
  \bibinfo {pages} {47} (\bibinfo {year} {1980})}\BibitemShut {NoStop}%
\bibitem [{\citenamefont {Haroche}\ and\ \citenamefont
  {Raimond}(2006)}]{haroche2006exploring}%
  \BibitemOpen
  \bibfield  {author} {\bibinfo {author} {\bibfnamefont {S.}~\bibnamefont
  {Haroche}}\ and\ \bibinfo {author} {\bibfnamefont {J.-M.}\ \bibnamefont
  {Raimond}},\ }\href@noop {} {\emph {\bibinfo {title} {Exploring the Quantum:
  Atoms, Cavities, and Photons}}}\ (\bibinfo  {publisher} {Oxford University
  Press},\ \bibinfo {year} {2006})\BibitemShut {NoStop}%
\bibitem [{\citenamefont {Scovil}\ and\ \citenamefont
  {Schulz-DuBois}(1959)}]{scovil1959three}%
  \BibitemOpen
  \bibfield  {author} {\bibinfo {author} {\bibfnamefont {H.~E.}\ \bibnamefont
  {Scovil}}\ and\ \bibinfo {author} {\bibfnamefont {E.~O.}\ \bibnamefont
  {Schulz-DuBois}},\ }\href@noop {} {\bibfield  {journal} {\bibinfo  {journal}
  {Phys. Rev. Lett.}\ }\textbf {\bibinfo {volume} {2}},\ \bibinfo {pages} {262}
  (\bibinfo {year} {1959})}\BibitemShut {NoStop}%
\bibitem [{\citenamefont {Klatzow}\ \emph {et~al.}(2019)\citenamefont
  {Klatzow}, \citenamefont {Becker}, \citenamefont {Ledingham}, \citenamefont
  {Weinzetl}, \citenamefont {Kaczmarek}, \citenamefont {Saunders},
  \citenamefont {Nunn}, \citenamefont {Walmsley}, \citenamefont {Uzdin},\ and\
  \citenamefont {Poem}}]{klatzow2019experimental}%
  \BibitemOpen
  \bibfield  {author} {\bibinfo {author} {\bibfnamefont {J.}~\bibnamefont
  {Klatzow}}, \bibinfo {author} {\bibfnamefont {J.~N.}\ \bibnamefont {Becker}},
  \bibinfo {author} {\bibfnamefont {P.~M.}\ \bibnamefont {Ledingham}}, \bibinfo
  {author} {\bibfnamefont {C.}~\bibnamefont {Weinzetl}}, \bibinfo {author}
  {\bibfnamefont {K.~T.}\ \bibnamefont {Kaczmarek}}, \bibinfo {author}
  {\bibfnamefont {D.~J.}\ \bibnamefont {Saunders}}, \bibinfo {author}
  {\bibfnamefont {J.}~\bibnamefont {Nunn}}, \bibinfo {author} {\bibfnamefont
  {I.~A.}\ \bibnamefont {Walmsley}}, \bibinfo {author} {\bibfnamefont
  {R.}~\bibnamefont {Uzdin}},\ and\ \bibinfo {author} {\bibfnamefont
  {E.}~\bibnamefont {Poem}},\ }\href@noop {} {\bibfield  {journal} {\bibinfo
  {journal} {Phys. Rev. Lett.}\ }\textbf {\bibinfo {volume} {122}},\ \bibinfo
  {pages} {110601} (\bibinfo {year} {2019})}\BibitemShut {NoStop}%
\bibitem [{\citenamefont {Mollier}\ \emph {et~al.}(1973)\citenamefont
  {Mollier}, \citenamefont {Hardin},\ and\ \citenamefont
  {Uebersfeld}}]{mollier73}%
  \BibitemOpen
  \bibfield  {author} {\bibinfo {author} {\bibfnamefont {J.~C.}\ \bibnamefont
  {Mollier}}, \bibinfo {author} {\bibfnamefont {J.}~\bibnamefont {Hardin}},\
  and\ \bibinfo {author} {\bibfnamefont {J.}~\bibnamefont {Uebersfeld}},\
  }\href@noop {} {\bibfield  {journal} {\bibinfo  {journal} {Rev. Sci.
  Instrum.}\ }\textbf {\bibinfo {volume} {44}},\ \bibinfo {pages} {1763}
  (\bibinfo {year} {1973})}\BibitemShut {NoStop}%
\bibitem [{\citenamefont {Henschel}\ \emph {et~al.}(2010)\citenamefont
  {Henschel}, \citenamefont {Majer}, \citenamefont {Schmiedmayer},\ and\
  \citenamefont {Ritsch}}]{henschel2010cavity}%
  \BibitemOpen
  \bibfield  {author} {\bibinfo {author} {\bibfnamefont {K.}~\bibnamefont
  {Henschel}}, \bibinfo {author} {\bibfnamefont {J.}~\bibnamefont {Majer}},
  \bibinfo {author} {\bibfnamefont {J.}~\bibnamefont {Schmiedmayer}},\ and\
  \bibinfo {author} {\bibfnamefont {H.}~\bibnamefont {Ritsch}},\ }\href@noop {}
  {\bibfield  {journal} {\bibinfo  {journal} {Phys. Rev. A}\ }\textbf {\bibinfo
  {volume} {82}},\ \bibinfo {pages} {033810} (\bibinfo {year}
  {2010})}\BibitemShut {NoStop}%
\bibitem [{\citenamefont {Oxborrow}(2007)}]{oxborrow2007}%
  \BibitemOpen
  \bibfield  {author} {\bibinfo {author} {\bibfnamefont {M.}~\bibnamefont
  {Oxborrow}},\ }\href@noop {} {\bibfield  {journal} {\bibinfo  {journal} {IEEE
  Trans. Micro.}\ }\textbf {\bibinfo {volume} {55}},\ \bibinfo {pages} {1209}
  (\bibinfo {year} {2007})}\BibitemShut {NoStop}%
\bibitem [{\citenamefont {Pollnau}\ \emph {et~al.}(1994)\citenamefont
  {Pollnau}, \citenamefont {Graf}, \citenamefont {Balmer}, \citenamefont
  {L{\"u}thy},\ and\ \citenamefont {Weber}}]{pollnau1994explanation}%
  \BibitemOpen
  \bibfield  {author} {\bibinfo {author} {\bibfnamefont {M.}~\bibnamefont
  {Pollnau}}, \bibinfo {author} {\bibfnamefont {T.}~\bibnamefont {Graf}},
  \bibinfo {author} {\bibfnamefont {J.}~\bibnamefont {Balmer}}, \bibinfo
  {author} {\bibfnamefont {W.}~\bibnamefont {L{\"u}thy}},\ and\ \bibinfo
  {author} {\bibfnamefont {H.}~\bibnamefont {Weber}},\ }\href@noop {}
  {\bibfield  {journal} {\bibinfo  {journal} {Phys. Rev. A}\ }\textbf {\bibinfo
  {volume} {49}},\ \bibinfo {pages} {3990} (\bibinfo {year}
  {1994})}\BibitemShut {NoStop}%
\bibitem [{\citenamefont {Patterson}\ \emph {et~al.}(1984)\citenamefont
  {Patterson}, \citenamefont {Lee}, \citenamefont {Wilson},\ and\ \citenamefont
  {Fayer}}]{patterson1984b}%
  \BibitemOpen
  \bibfield  {author} {\bibinfo {author} {\bibfnamefont {F.~G.}\ \bibnamefont
  {Patterson}}, \bibinfo {author} {\bibfnamefont {H.~W.~H.}\ \bibnamefont
  {Lee}}, \bibinfo {author} {\bibfnamefont {W.~L.}\ \bibnamefont {Wilson}},\
  and\ \bibinfo {author} {\bibfnamefont {M.~D.}\ \bibnamefont {Fayer}},\
  }\href@noop {} {\bibfield  {journal} {\bibinfo  {journal} {Chem. Phys.}\
  }\textbf {\bibinfo {volume} {84}},\ \bibinfo {pages} {51} (\bibinfo {year}
  {1984})}\BibitemShut {NoStop}%
\bibitem [{\citenamefont {Takeda}\ \emph {et~al.}(2002)\citenamefont {Takeda},
  \citenamefont {Takegoshi},\ and\ \citenamefont {Terao}}]{takeda2002zero}%
  \BibitemOpen
  \bibfield  {author} {\bibinfo {author} {\bibfnamefont {K.}~\bibnamefont
  {Takeda}}, \bibinfo {author} {\bibfnamefont {K.}~\bibnamefont {Takegoshi}},\
  and\ \bibinfo {author} {\bibfnamefont {T.}~\bibnamefont {Terao}},\
  }\href@noop {} {\bibfield  {journal} {\bibinfo  {journal} {J. Chem. Phys.}\
  }\textbf {\bibinfo {volume} {117}},\ \bibinfo {pages} {4940} (\bibinfo {year}
  {2002})}\BibitemShut {NoStop}%
\bibitem [{\citenamefont {Nelson}\ \emph {et~al.}(1981)\citenamefont {Nelson},
  \citenamefont {Lutz}, \citenamefont {Fayer},\ and\ \citenamefont
  {Madison}}]{nelson1981laser}%
  \BibitemOpen
  \bibfield  {author} {\bibinfo {author} {\bibfnamefont {K.~A.}\ \bibnamefont
  {Nelson}}, \bibinfo {author} {\bibfnamefont {D.}~\bibnamefont {Lutz}},
  \bibinfo {author} {\bibfnamefont {M.}~\bibnamefont {Fayer}},\ and\ \bibinfo
  {author} {\bibfnamefont {L.}~\bibnamefont {Madison}},\ }\href@noop {}
  {\bibfield  {journal} {\bibinfo  {journal} {Phys. Rev. B}\ }\textbf {\bibinfo
  {volume} {24}},\ \bibinfo {pages} {3261} (\bibinfo {year}
  {1981})}\BibitemShut {NoStop}%
\bibitem [{\citenamefont {Deeg}\ \emph {et~al.}(1985)\citenamefont {Deeg},
  \citenamefont {Madison},\ and\ \citenamefont {Fayer}}]{deeg1985}%
  \BibitemOpen
  \bibfield  {author} {\bibinfo {author} {\bibfnamefont {F.~W.}\ \bibnamefont
  {Deeg}}, \bibinfo {author} {\bibfnamefont {L.}~\bibnamefont {Madison}},\ and\
  \bibinfo {author} {\bibfnamefont {M.}~\bibnamefont {Fayer}},\ }\href@noop {}
  {\bibfield  {journal} {\bibinfo  {journal} {Chem. Phys.}\ }\textbf {\bibinfo
  {volume} {94}},\ \bibinfo {pages} {265} (\bibinfo {year} {1985})}\BibitemShut
  {NoStop}%
\bibitem [{\citenamefont {Yang}\ \emph {et~al.}(2000)\citenamefont {Yang},
  \citenamefont {Sloop}, \citenamefont {Weissman},\ and\ \citenamefont
  {Lin}}]{yang2000zero}%
  \BibitemOpen
  \bibfield  {author} {\bibinfo {author} {\bibfnamefont {T.-C.}\ \bibnamefont
  {Yang}}, \bibinfo {author} {\bibfnamefont {D.~J.}\ \bibnamefont {Sloop}},
  \bibinfo {author} {\bibfnamefont {S.}~\bibnamefont {Weissman}},\ and\
  \bibinfo {author} {\bibfnamefont {T.-S.}\ \bibnamefont {Lin}},\ }\href@noop
  {} {\bibfield  {journal} {\bibinfo  {journal} {J. Chem. Phys.}\ }\textbf
  {\bibinfo {volume} {113}},\ \bibinfo {pages} {11194} (\bibinfo {year}
  {2000})}\BibitemShut {NoStop}%
\bibitem [{\citenamefont {Penfield}(1962)}]{penfield1962noisewave}%
  \BibitemOpen
  \bibfield  {author} {\bibinfo {author} {\bibfnamefont {P.}~\bibnamefont
  {Penfield}},\ }\href@noop {} {\bibfield  {journal} {\bibinfo  {journal} {IRE
  Trans. Circuit Theory}\ }\textbf {\bibinfo {volume} {\textit{March}}},\
  \bibinfo {pages} {84} (\bibinfo {year} {1962})}\BibitemShut {NoStop}%
\bibitem [{\citenamefont {\textrm{PMA2-33LN+ datasheet and .s2p
  file}}()}]{PMA2}%
  \BibitemOpen
  \bibfield  {author} {\bibinfo {author} {\bibnamefont {\textrm{PMA2-33LN+
  datasheet and .s2p file}}},\ }\href@noop {} {}\bibinfo {address}
  {Mini-Circuits (2019)}\BibitemShut {NoStop}%
\bibitem [{\citenamefont {Kraus}(1986)}]{kraus1986radio}%
  \BibitemOpen
  \bibfield  {author} {\bibinfo {author} {\bibfnamefont {J.~D.}\ \bibnamefont
  {Kraus}},\ }\href@noop {} {\emph {\bibinfo {title} {Radio astronomy}}}\
  (\bibinfo  {publisher} {Cygnus-Quasar Books, Powell, OH},\ \bibinfo {year}
  {1986})\BibitemShut {NoStop}%
\bibitem [{\citenamefont {Oxborrow}(2017)}]{oxborrow2017maser}%
  \BibitemOpen
  \bibfield  {author} {\bibinfo {author} {\bibfnamefont {M.}~\bibnamefont
  {Oxborrow}},\ }\href@noop {} {\bibinfo {title} {Maser assembly}} (\bibinfo
  {year} {2017}),\ \bibinfo {note} {{US} Patent 9,608,396}\BibitemShut
  {NoStop}%
\end{thebibliography}

\providecommand{\noopsort}[1]{}\providecommand{\singleletter}[1]{#1}%

\end{document}

\preprint{APS/123-QED}

\title{Supplementary Material:\\ Bench-top Cooling of a Microwave Mode using an Optically Pumped Spin Refrigerator}% Force line breaks with \\
%\thanks{A footnote to the article title}%https://www.overleaf.com/project/5e1dfb30feae1900017dea0f

\author{Hao Wu, Shamil Mirkhanov, Wern Ng}
\author{Mark Oxborrow}%
 \email{m.oxborrow@imperial.ac.uk}
\affiliation{%
Department of Materials, Imperial College London, Exhibition Road, London SW7 2AZ, United Kingdom
}%


%\affiliation{
%\affiliation{
 %Third institution, the second for Charlie Author
%}%
%author{Delta Author}
%\affiliation{%



%\begin{abstract}
%An article usually includes an abstract, a concise summary of the work
%covered at length in the main body of the article. 
%\begin{description}
%\end{description}
%\end{abstract}

\maketitle

%\tableofcontents

\section{\label{sec:exp_det1}Experimental Method \--Details}
At its operating frequency of $f_\text{mode} \equiv \omega_\text{mode} / 2 \pi =$~1.4495~GHz,
the TE$_{01\delta}$ mode of our microwave cavity had a loaded quality factor of $Q_\textrm{L} \approx 3600$ at critical coupling,
equating to an unloaded quality factor of $Q_0 \approx 2 Q_\textrm{L} = 7200$.
The cavity's internal cylindrical dielectric ring was made of monocrystalline strontium-titanate (STO), 8.5~mm in height with outer and inner diameters of 12 and 4~mm, respectively. All of the ring's surfaces were finely polished without sleeks. The copper cavity that concentrically surrounded the ring was constructed through ham-radio ``plumbing" techniques: its cylindrical side wall and roof were made from a standard copper pipe fitting, viz.~a 28-mm end-feed end cap. The cavity's internal height ($\approx$~15~mm) could be adjusted using an internal piston in the form of a 26-mm dia.~copper disk suspended by a brass screw, itself held by a brass hex nut soldered onto a centered hole in the end cap's roof. The cavity's base was a piece of copper-coated printed circuit board (PCB), to which the end cap was clamped.  A post made of cross-linked polystyrene (an equivalent of Rexolite), mounted into a hole in the PCB, held the STO ring $\sim\!3$~mm above the PCB and concentrically with respect to the cavity's side wall (viz. the end cap's barrel). All working copper surfaces were polished with Brasso then rubbed clean with acetone. Through a finite-element-method (FEM) electromagnetic simulation using COMSOL Multiphysics\cite{oxborrow2007,wu2020room}, the magnetic mode volume $V_\textrm{mode}$ of the resonator's TE$_{01\delta}$ mode was estimated to be $\approx0.32$~cm$^3$. 

A monocrystal of pentacene-doped \textit{p-}terphenyl was shaped and slotted snugly into the STO ring's 4-mm bore (though did not entirely fill it). The long-pulsed 590-nm dye laser used for our cooling experiments provided a collimated beam $\sim\!2$~mm in diameter, which entered through a 3-mm hole in the cavity's solid copper wall. Each shot from the laser comprised a train of three pulses, each 150~$\mu$s in duration and staggered at intervals of 500~$\mu$s; the total energy per shot (for the data shown in Fig.~4) was 2.4~J, as measured with a calibrated bolometer.
The geometry and ratios of refractive indices were such that the STO ring functioned as a cylindrical beam compressor:
the interface between the STO's outer cylindrical surface and the air around it functioned as a cylindrical converging lens reducing the width of the pump beam by a factor of $\sim\!$~0.61. The cylindrical interface between the STO and crystal functioned as a compensating  cylindrical diverging lens such that the pump beam propagating through the \textit{p-}terphenyl crystal was near-collimated, its elliptical cross-section being $\sim\!2$~mm in height (major diameter) and $\sim\!1.2$~mm in width (minor diameter).

The volume of pentacene-doped \textit{p}-terphenyl illuminated by the pump beam was thus, approximately, an elliptical prism of the above cross-section and $\leq4\!$~mm in length/depth through the crystal. The centers of the prism's two (cylindrically rounded) end faces lay at opposite azimuthal locations on the equatorial circle, half way up the inner cylindrical wall of the STO ring, where the magnitude of the TE$_{01\delta}$ mode's a.c.~magnetic flux density, $\textbf{B}$, is maximum. The magnitude of $\textbf{B}$ throughout the illuminated prism was found to be not less than 90\% of the mode's maximum value. The direction of magnetic flux density vector $\textbf{B}$ throughout the prism was found to deviate by no more than 5\degree$\,$ from the cylindrical axis. This relative uniformity of $\textbf{B}$ justifies the invocation of a mean-field approximation (detailed in the next section), where all photo-excited pentacene molecules see the same driving a.c. magnetic field.

Our heterodyne receiver [lower section of Fig.~3(a)] was similar in concept to that used for the detection of weak maser emissions from  Rydberg atoms\cite{moi1980heterodyne}. Sufficient sensitivity was achieved through  a cascade of amplifiers (total gain $\sim\!100$~dB) whose output was terminated through a narrow-band-pass ``roofing'' filter into a logarithmic detector possessing a dynamic range of 80 dB. This filter was of quartz-SAW type, operating at a central frequency (C.F.) of 70 MHz, with a steep-skirted bandwidth (B.W.) of $B_{\scriptscriptstyle\text{SAW}}=$~50 kHz. The reception bandwidth of the receiver thus fitted easily within the linewidth of the cavity's TE$_{01\delta}$ mode \--see Fig.~3(b). All components operated under ambient lab conditions. Note in particular that, to improve sensitivity (SNR) at the expense of extra calibrational complexity, no (room-temperature) isolator was inserted between the cavity's out-coupling loop and the receiver's front end LNA.

\section{\label{sec:level1}Model of Pentacene's Spin Dynamics}

\begin{figure}[b!]
\includegraphics[scale=0.60]{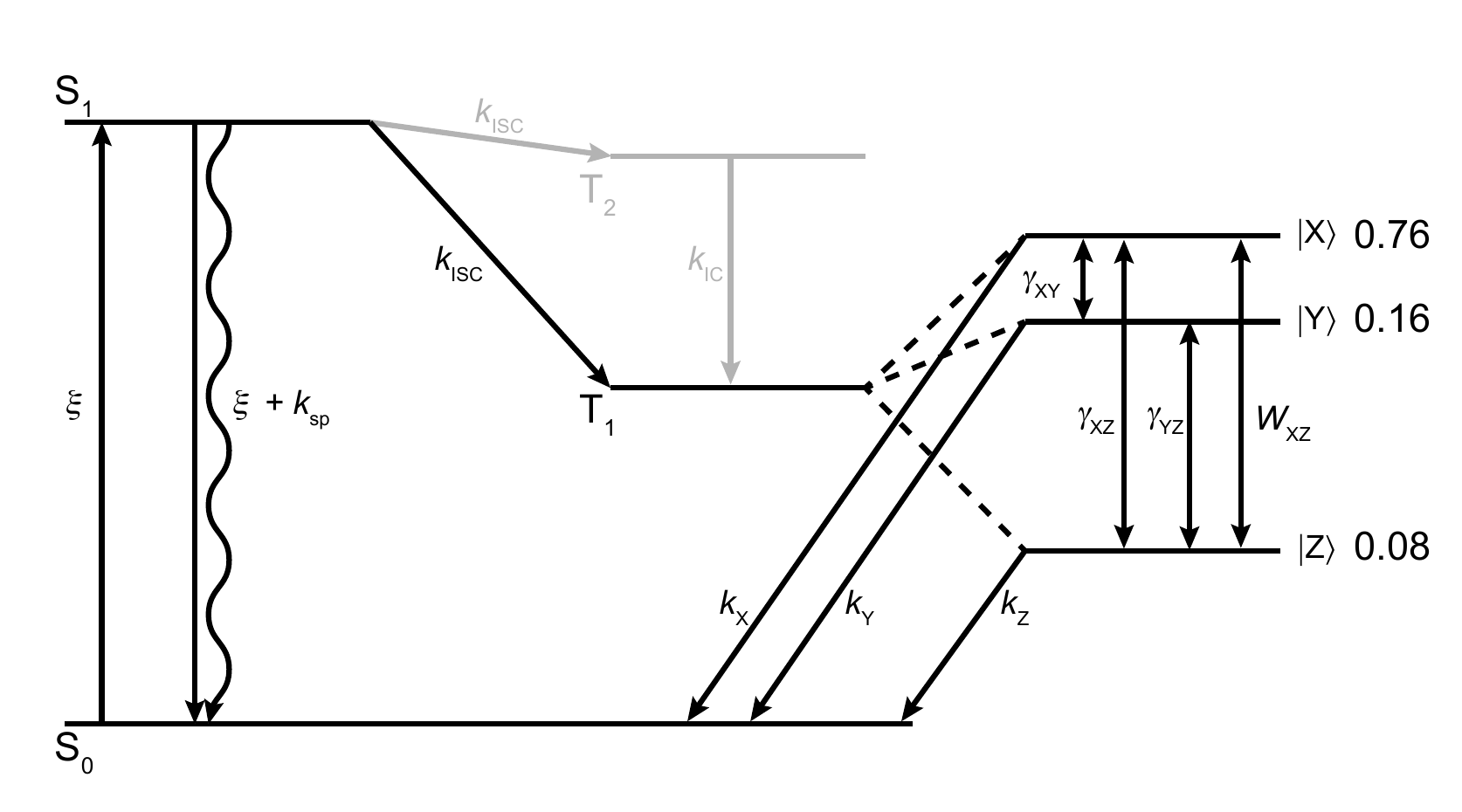}% Here is how to import EPS art
\caption{\label{fig:S1} Zero-field spin dynamics of molecular pentacene.}
\end{figure}

The model we used to simulate the spin dynamics of molecular pentacene is shown schematically in Fig.~\ref{fig:S1}.
The corresponding rate equations, here shown in matrix form, were integrated to calculate the time-dependent population in each state/sub-level. 
\begin{equation}
\label{eq:eqS1}
\resizebox{0.93\hsize}{!}{% this option is for resizing
    $\begin{bmatrix}
   \dot{S_0}\\\dot{S_1}\\\dot{N_\textrm{X}}\\\dot{N_\textrm{Y}}\\\dot{N_\textrm{Z}}\\
    \end{bmatrix}
    =
    \begin{bmatrix}
     -\xi & \xi+k_\textrm{sp} & k_\textrm{X} & k_\textrm{Y} & k_\textrm{Z}\\
     \xi & -(\xi+k_\textrm{sp}+k_\textrm{ISC}) & 0 & 0 & 0\\
    0 & P_\textrm{X}k_\textrm{ISC} & -(k_\textrm{X}+\gamma_\textrm{XY}+\gamma_\textrm{XZ}+W_\textrm{XZ}) & \gamma_\textrm{XY} & \gamma_\textrm{XZ}+W_\textrm{XZ}\\
    0 & P_\textrm{Y}k_\textrm{ISC} & \gamma_\textrm{XY} & -(k_\textrm{Y}+\gamma_\textrm{XY}+\gamma_\textrm{YZ}) & \gamma_\textrm{YZ}\\
     0 & P_\textrm{Z}k_\textrm{ISC} & \gamma_\textrm{XZ}+W_\textrm{XZ} & \gamma_\textrm{YZ} & -(k_\textrm{Z}+\gamma_\textrm{XZ}+\gamma_\textrm{YZ}+W_\textrm{XZ})\\
    \end{bmatrix}
    \begin{bmatrix}
    S_0\\S_1\\N_\textrm{X}\\N_\textrm{Y}\\N_\textrm{Z}
    \end{bmatrix}$
}
\end{equation}
Here,  $S_0$, $S_1$, $N_\textrm{X}$, $N_\textrm{Y}$ and $N_\textrm{Z}$ denote populations in their associated states/sub-levels as shown in Fig.~\ref{fig:S1};
$\xi$ is the optical pumping parameter\cite{pollnau1994explanation};
$W_\textrm{XZ}$ are the rates of stimulated transitions; definitions of the remaining symbols are given in
Table.~\ref{tab:tableS1} below.
The values stated in this table's right-hand column are best-estimates for the room-temperature properties of pentacene doped (at $\sim\!0.1$~\% conc.) into \textit{p}-terphenyl.
\begin{table}[htbp!]%The best place to locate the table environment is directly after its first reference in text
\caption{\label{tab:tableS1}%
Summarized definitions and values of the symbols used in the model
}
\centering
\scalebox{1.0}{
%\begin{ruledtabular}
\begin{tabular}{llr}
\colrule
\colrule
\textrm{Symbol}& 
\textrm{Definition}&
\multicolumn{1}{c}{\textrm{Value}}\\[0pt]
\colrule
& & \\[-8pt]
$k_\textrm{sp}$ & rate of spontaneous emission from \text{S}$_1$ $\rightarrow$ \text{S}$_0$  & 4.2$\times10^7$  s$^{-1}$\cite{patterson1984b,takeda2002zero}\\[1pt]
$k_\textrm{ISC}$ & rate of intersystem crossing from \text{S}$_1$ $\rightarrow$ \text{T}$_1$ & 6.9$\times10^7$  s$^{-1}$\cite{patterson1984b,takeda2002zero}\\[1pt]
$P_\textrm{X}:P_\textrm{Y}:P_\textrm{Z}\;\;\;$ & normalized population ratios into T$_1$'s sub-levels & $\;\;\;0.76:0.16:0.08$\cite{sloop1981electron}\\[1pt]
$\gamma_\textrm{XY}$ & rate of spin-lattice relaxation between $|\text{X}\rangle$ and $|\text{Y}\rangle$ & 0.4$\times10^4$ s$^{-1}$\cite{wu2019unraveling}\\[1pt]
$\gamma_\textrm{YZ}$ & rate of spin-lattice relaxation between $|\text{Y}\rangle$ and $|\text{Z}\rangle$ & 2.2$\times10^4$ s$^{-1}$\cite{wu2019unraveling}\\[1pt]
$\gamma_\textrm{XZ}$ & rate of spin-lattice relaxation between $|\text{X}\rangle$ and $|\text{Z}\rangle$ & 1.1$\times10^4$ s$^{-1}$\cite{wu2019unraveling}\\[1pt]
$k_\textrm{X}$ & decay rate of $|\text{X}\rangle \rightarrow$ \text{S}$_0$ & 2.8$\times10^4$ s$^{-1}$\cite{wu2019unraveling}\\[1pt]
$k_\textrm{Y}$ & decay rate of $|\text{Y}\rangle \rightarrow$ \text{S}$_0$ & 0.6$\times10^4$ s$^{-1}$\cite{wu2019unraveling}\\[1pt]
$k_\textrm{Z}$ & decay rate of $|\text{Z}\rangle \rightarrow$ \text{S}$_0$ & 0.2$\times10^4$ s$^{-1}$\cite{wu2019unraveling}\\[1pt]
\colrule
\end{tabular}
}
%\end{ruledtabular}
\end{table}
The rapidity and polarization-preserving nature of the internal conversion from $\textrm{T}_2$ to $\textrm{T}_1$ (grayed section of Fig.~\ref{fig:S1})
allows the dynamics to be simplified such that population moves from $\textrm{S}_1$ directly into the three sub-levels of T$_1$ at rates equal to the overall ISC rate, $k_\textrm{ISC}$, multiplied by the normalized population (or ``splitting'') ratio for each sub-level.

The pumping parameter $\xi$ can be expressed as\cite{pollnau1994explanation} 
\begin{equation}
    \xi =  \frac{\lambda_\text{p}}{h c l A_\text{p}}P(t)\{1-\text{exp}[-l\alpha]\}\frac{\sigma_\textrm{a}}{\alpha},
\label{eq:eqS2}
\end{equation}
where $\lambda_\text{p}$ is the pump's wavelength (590~nm), $h$ Planck's constant, $c$ the speed of light in a vacuum,
$l$ the thickness of the crystal ($\leq 4$~mm) as traversed by the pump beam,
$A_\text{p}$ the cross-sectional area ($\sim\!1.9$~mm$^2$) of same,
$P(t)$ the instantaneous power of same,
$\sigma_\textrm{a}$ the absorption cross section of molecular pentacene when doped in \textit{p}-terphenyl ($\approx~2\times10^{-17}$~cm$^2$\cite{nelson1981laser}) at $\lambda_\text{p}$,
and $\alpha$ the crystal's optical absorption coefficient at $\lambda_\text{p}$.
For a sufficiently powerful pump beam, the crystal becomes bleached\cite{deeg1985,takeda2002zero} and optically thin:
$1-\text{exp}[-l\alpha] \approx l\alpha$. Eq.~(\ref{eq:eqS2}) thereupon simplifies to
\begin{equation}
    \xi \approx \frac{\lambda_\textrm{p} \sigma_\textrm{a}}{hc A_\text{p}}P(t).
\label{eq:eqS3}
\end{equation}
\begin{figure}[b!]
\includegraphics{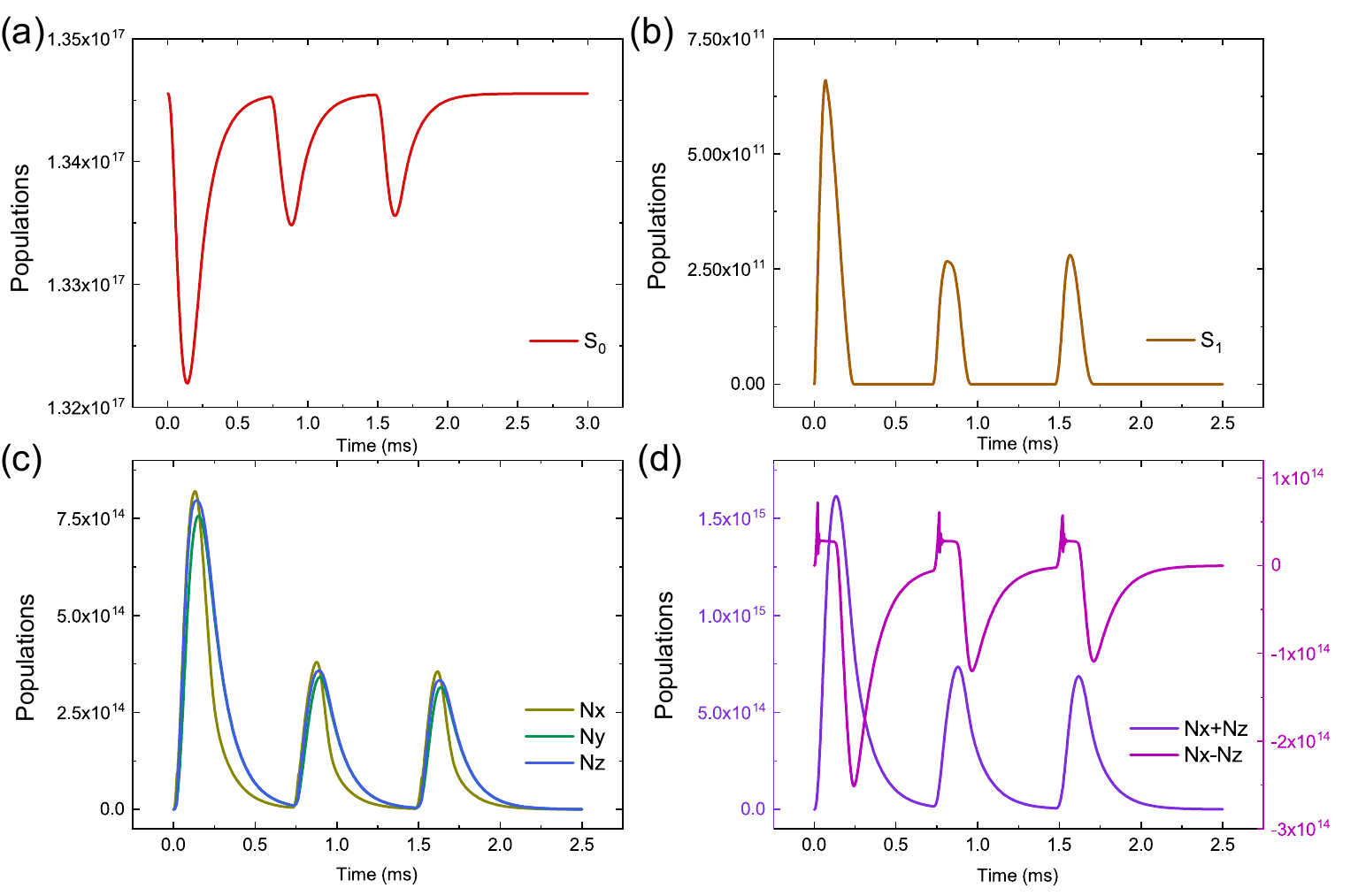}% Here is how to import EPS art
\caption{\label{fig:S2} Simulated dynamics of the populations in (a) pentacene's singlet ground state S$_0$, (b) its first excited singlet state S$_1$ and (c) the three sub-levels of its lowest triplet state T$_1$. (d) The sum and difference of the populations in the $|\text{X}\rangle$ and $|\text{Z}\rangle$ sub-levels of T$_1$.}
\end{figure}
The cavity's TE$_{01\delta}$ mode holds electromagnetic energy, $W$, that is smoothly juggled (preserving $W$) between standing electric and magnetic fields at
a frequency of $f_\text{mode}$.
This energy is dissipated through both dielectric losses in the STO ring and resistive heating
in the copper surfaces that surround the TE$_{01\delta}$ mode.
In units of power, the rate of dissipation equals
 $P_0 = \omega_\text{mode} W  Q^{-1}_\textrm{0}$, where $\omega_\text{mode} = 2\pi f_\textrm{mode}$ and
 $Q_\textrm{0}$ is the TE$_{01\delta}$ mode's intrinsic (= ``unloaded'') quality factor.
 If the cavity is connected to an external load, like the input of a low-noise amplifier (LNA),
 this load will also sink energy from the mode,
 its dissipation rate being  $P_\textrm{ex}  = \omega_\text{mode} W  Q^{-1}_\textrm{ex}$, where $Q_\textrm{ex}$ is the so-called external quality factor. Here, under critical coupling, $Q_0=Q_\textrm{ex}=7200$.
Upon optical pumping, the a.c. magnetic field of the cavity's TE$_{01\delta}$ mode further interacts with pentacene molecules in the
 $\ket{\text{X}}$ and $\ket{\text{Z}}$ sub-levels of pentacene's T$_1$ state.
Preserving ref.~\onlinecite{sieg1964microwave}'s sign convention (despite its awkwardness for analysing masar cooling),
the rate at which energy is \textit{added} to the mode equals $\omega_\text{mode} W / Q_\textrm{m}^{-1}$,
where  $Q_\textrm{m}$ is the mode's so-called magnetic quality factor\cite{sieg1964microwave}. 
The \textit{removal} of energy from the mode by masar cooling thus corresponds to a \textit{negative} $Q_\textrm{m}$.
The stimulated transition rate between $|\text{X}\rangle$ and $|\text{Z}\rangle$, $W_\textrm{XZ}$ obeys\cite{salvadori2017nanosecond}
\begin{equation}
    W_\textrm{XZ} = Bq,
\end{equation}
where $B$ is Einstein's \textit{B} coefficient for the $|\text{X}\rangle \leftrightarrow |\text{Z}\rangle$ transition and $q$ is the number of photons in the TE$_{01\delta}$ mode.
The former is evaluated through\cite{sieg1964microwave}
\begin{equation}
    B  =  \frac{\mu_0 \gamma^2 h f_\text{mode} T_2 \braket{\sigma^2}}{2V_\textrm{mode}},
     \label{eq:EinsteinB}
\end{equation}
where $\mu_0$ is the permeability of free-space;
$\gamma \approx 2 \pi \times 28$~GHz/T is the (reciprocal) gyromagnetic ratio for the X-Z transition; 
$T_2\sim\!2.9$ $\mu$s is the spin-spin dephasing time associated with the X-Z transition\cite{yang2000zero};
the normalized transition probability matrix element $\braket{\sigma^2}$  equals $0.5$ for a transition between two sub-levels of a triplet ($S = 1$) state, where
the transition is stimulated by a microwave mode with a linearly polarized a.c.~magnetic field \--see Table 5-1 in ref.~\onlinecite{sieg1964microwave};
$B$ evaluates to  $\sim\!9\times10^{-8}$ s$^{-1}$.
The number of photons $q$ in the mode obeys the rate equation
\begin{equation}
    \dot{q} = - \omega_\text{mode} [ (Q_0^{-1}+Q_\textrm{ex}^{-1}) (q - \epsilon T_0)] +
    B (N_\textrm{X}-N_\textrm{Z}) (q - \epsilon  T_\text{X-Z}),
    \label{eq:qdot}
\end{equation}
where the constant $\epsilon = k_{B}/( h f_\text{mode})$ converts absolute temperature into the expected number of photons for a microwave mode (regarded as an oscillator with two independent degrees of freedom) of frequency $f_\text{mode}$ obeying Maxwell-Boltzmann equipartition;
$k_{B}$ is Boltzmann's constant;
$T_0 = 290$~K is the temperature of the cavity's lossy (dielectric and metallic) materials and the receiver's 50-$\ohm$ input as our experiments were carried out at room temperature;
$T_\text{X-Z}$ is the temperature of the internal magnetic = ``spin''  system (viz.~the X-Z transition) whose value is extremely small ($\sim$ mK) during masing and masar cooling actions (as proven in the main text) and can be safely neglected in the calculation. 
$N_\textrm{X}$ and $N_\textrm{Z}$ are the populations of  the $|\text{X}\rangle$ and $|\text{Z}\rangle$ sub-levels, respectively, as obeyed by Eq.~(\ref{eq:eqS1}).

The combined rate equations [viz.~Eqs.~(\ref{eq:eqS1}) and (\ref{eq:qdot})]
were solved by the Runge-Kutta method for the experimentally measured instantaneous pump power $P(t)$ [orange trace in Fig.~4(a)].
The resultant time-integrated populations are shown in Fig.~\ref{fig:S2} and the time evolution of $q$ is demonstrated in Fig.~4(b) and (d) in the forms of $T_\textrm{mode}$ and $P_\textrm{maser}$, respectively (see main text).

\section{Noise Analysis}
 \begin{figure}[b!]
\includegraphics{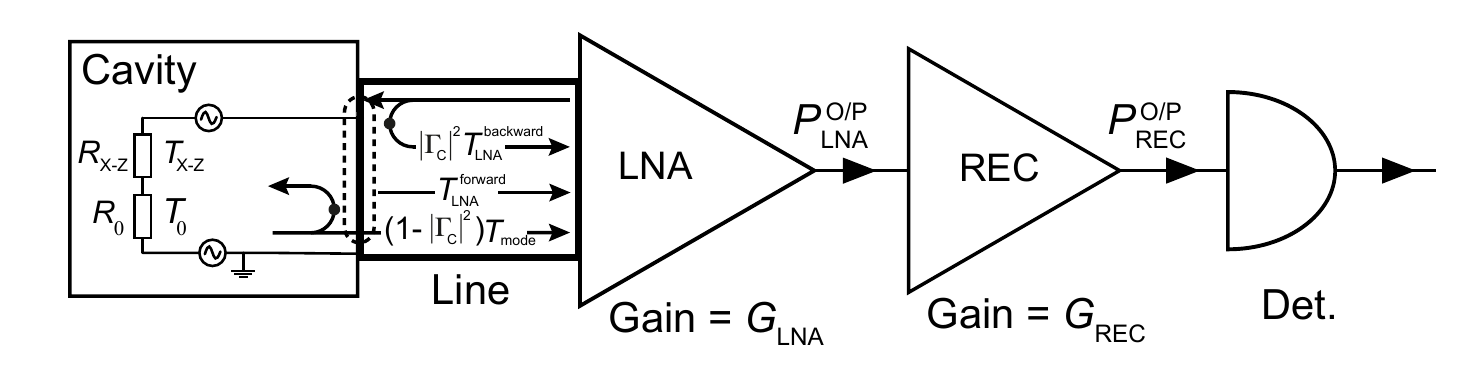}% Here is how to import EPS art
\caption{\label{fig:S3}Anatomy and noise wave representation, focusing on the heterodyne receiver's front end.}
\end{figure}

We here adopt the ``wave approach''\cite{penfield1962noisewave,sieg1964microwave,djordjevic2017wave} for
analysing the flow of noise power from the cavity into our heterodyne receiver as shown in 
Fig.~\ref{fig:S3}.
We shall first focus on the front-end of the receiver where the cavity and the first low-noise amplifier (LNA) are connected by a 4-cm length of 50-$\ohm$ low-loss semi-rigid (RG405) coaxial cable. This cable's shortness allows both its phase length and the thermal (Johnson) noise that it generates to be ignored to adequate first approximation, so simplifying the noise analysis. 

The noise at the LNA's input must include the amplifier's forward-propagating noise wave together with that fraction of its backward-propagating noise wave that gets reflected off the cavity into the forward direction, together with the ``interference'' between these two (partially correlated) waves. 
The input noise also includes the fraction of the cavity's noise that escapes through the coupling port. It is convenient to quantify each noise power per unit bandwidth (divided by $k_\textrm{B}$) as a temperature in units of K; see Fig.~\ref{fig:S3}.
%the above two mentioned noise waves %associated with the  LNA, whose net %contribution is quantified by the %equivalent temperature $T_\textrm{LNA}$, %together with noise waves from the %cavity's and from the (lossy) joining  %coaxial cable, whose net noise power are %quantified by the equivalent temperatures %$T_\textrm{mode}$ and $T_\textrm{cable}$, %respectively
According to ref.~\onlinecite{djordjevic2017wave}, the noise temperature referred to the LNA's input $T_\textrm{ns}$ equals
\begin{equation}\label{eq:S7}
    T_\textrm{ns} = (T_\textrm{mode} +  T_\textrm{LNA})(1-|\Gamma_\textrm{c}|^2),
\end{equation}
where $T_\textrm{mode}$
is the ``source'' noise of the cavity mode and 
\begin{equation}\label{eq:S8}
T_\textrm{LNA} = T_\textrm{min}+4T_0\frac{R_\textrm{n}}{Z_0}\frac{|\Gamma_\textrm{c}-\Gamma_\textrm{opt}|^2}{|1+\Gamma_\textrm{opt}|^2(1-|\Gamma_\textrm{c}|^2)}.
\end{equation}
Eq.~(\ref{eq:S8}) is the industry-standard formula and parameterization
for quantifying an LNA's noise behavior, 
 where estimates of the parameters appearing within it are provided in the LNA's datasheet as supplied by the manufacturer. For the particular LNA that we used (as the front-end amplifier of our heterodyne receiver)\cite{PMA2}, the minimum noise temperature $T_\textrm{min}=17.4$ K;
the ratio of the noise resistance to the ($Z_0 = 50~\Omega$) normalization impedance $\frac{R_\textrm{n}}{Z_0}=0.022$;
and the magnitude and angle of the optimum source reflection coefficient $\Gamma_\textrm{opt}=-0.131+0.189i$.
What makes the noise calibration of our heterodyne receiver (lacking a front-end isolator) tricky is that the reflection coefficient of the cavity (as seen by the LNA's input), $\Gamma_\textrm{c}$, is itself dependent on the TE$_{01\delta}$ mode's magnetic quality factor,
$Q_\textrm{m}$ (as controls the impedance $R_\text{X-Z}$ appearing in Fig.~\ref{fig:S3}),
which in turn controls the mode temperature $T_\textrm{mode}$, where both $Q_\textrm{m}$ (and so $R_\text{X-Z}$) and $T_\textrm{mode}$ depend on time. 
These relations are worked out below.
%In Eq.~\ref{eq:S9}, $T_\textrm{cable} = %(L_\textrm{dBcable}/4.34)T_0 = 2.1$ %K\cite{sieg1964microwave} %is the noise %temperature of the 4-cm-long length of %coaxial %cable connecting the cavity and %the first LNA, assuming a %uniform %temperature along the cable of $T_0=290$ %K, where %$L_\textrm{dBcable}=0.032$ dB is %the loss of the cable.

In addition to the noise sources mentioned above, the image noise [associated with downconverting the signal to a 70-MHz ``IF'' using a standard (i.e., not SSB) mixer] at 1.3095 GHz also contributes to the noise power (referenced to the first LNA's input), whose noise temperature has the same form as Eq.~\ref{eq:S7}. To evaluate the image noise temperature, the first LNA can be considered as being connected to the same cavity but off-resonance, where the coupling loop acts as a short and thus $\Gamma_\textrm{c}=-1$. Combining Eqs.~\ref{eq:S7} and \ref{eq:S8}, the image noise temperature $T_\textrm{image}$ can be expressed as
\begin{equation}
    T_\textrm{image} = 4T_0\frac{R_\textrm{n}}{Z_0}\frac{|-1-\Gamma_\textrm{opt}|^2}{|1+\Gamma_\textrm{opt}|^2}.
\end{equation}
For the noise-parameter values stated above, $T_\textrm{image}=25.5$ K.
Therefore, the sum of noise temperatures referenced to the first LNA's input, 
\begin{equation}\label{eq:S11}
    T_\textrm{front-end} = (T_\textrm{LNA}+T_\textrm{mode})(1-|\Gamma_\textrm{c}|^2)+T_\textrm{image},
\end{equation}
and the output of the first LNA 
\begin{equation}\label{eq:S12}
P_\textrm{LNA}^\textrm{O/P} = G_\textrm{LNA}k_\textrm{B}B_\textrm{SAW}T_\textrm{front-end},
\end{equation}
where the power gain of the first LNA $G_\textrm{LNA} = 32.5$ (expressed in dimensionless linear units) and $B_\textrm{SAW}=50$ kHz is the bandwidth of the heterodyne receiver.

The noise factor of the rest of the receiver, excluding the front-end LNA, equals $F_\textrm{REC}=1.15$; this  is determined (to adequate first approximation) by the Friis formula\cite{kraus1986radio}. Hence, the noise temperature of the rest of the receiver $T_\textrm{REC}=(F_\textrm{REC}-1)T_0=43$ K, referenced to the industry standard reference temperature $T_0$ of 290 K. Combining Eq.~\ref{eq:S8} and Eq.~\ref{eq:S11} to \ref{eq:S12}, the heterodyne receiver's output power can thus be expressed as
\begin{equation}\label{eq:S13}
\begin{split}
    P_\textrm{REC}^\textrm{O/P} & = G_\textrm{REC}(P_\textrm{LNA}^\textrm{O/P}+k_\textrm{B}B_\textrm{SAW}T_\textrm{REC})\\
    & = G_\textrm{REC}k_\textrm{B}B_\textrm{SAW}\{G_\textrm{LNA}[(T_\textrm{min}+T_\textrm{mode})(1-|\Gamma_\textrm{c}|^2)+4T_0\frac{R_\textrm{n}}{Z_0}\frac{|\Gamma_\textrm{c}-\Gamma_\textrm{opt}|^2}{|1+\Gamma_\textrm{opt}|^2}+T_\textrm{image}]+T_\textrm{REC}\}
\end{split}
\end{equation}
where $G_\textrm{REC}$ is the total gain of the rest of the receiver.

According to Eq.~\ref{eq:S13}, under masar cooling, the power reduction at the heterodyne receiver's output can be expressed in units of dB as:
\begin{equation}\label{eq:S14}
\begin{split}
    \Delta P &= 10\log_{10}\frac{P_\textrm{REC}^\textrm{O/P}}{P_\textrm{REC}^\textrm{O/P(0)}}\\
    & = 10\log_{10}\frac{G_\textrm{LNA}[(T_\textrm{min}+T_\textrm{mode})(1-|\Gamma_\textrm{c}|^2)+4T_0\frac{R_\textrm{n}}{Z_0}\frac{|\Gamma_\textrm{c}-\Gamma_\textrm{opt}|^2}{|1+\Gamma_\textrm{opt}|^2}+T_\textrm{image}]+T_\textrm{REC}}{G_\textrm{LNA}[(T_\textrm{min}+T_\textrm{mode}^0)(1-|\Gamma_\textrm{c}^0|^2)+4T_0\frac{R_\textrm{n}}{Z_0}\frac{|\Gamma_\textrm{c}^0-\Gamma_\textrm{opt}|^2}{|1+\Gamma_\textrm{opt}|^2}+T_\textrm{image}]+T_\textrm{REC}}
\end{split}
\end{equation}
where $P_\textrm{REC}^\textrm{O/P(0)}$ is the receiver's output power when the cavity is not pumped and $T_\textrm{mode}^0=290$ K and $\Gamma_\textrm{c}^0=0$ are the initial conditions of the microwave mode's noise temperature and the reflection coefficient of the cavity, respectively. Thus, we have now obtained the dependence of $\Delta P$ on  $T_\textrm{mode}$ and $\Gamma_\textrm{c}$. The next step is to find out the dependence of $\Gamma_\textrm{c}$ on $T_\textrm{mode}$, which is worked out below.

Upon optical pumping, spin polarized molecules are generated inside the cavity, thus changing
the cavity's impedance to $Z_{\scriptscriptstyle\text{CAV}} =  R_0 +R_\text{X-Z}$, see Fig.~\ref{fig:S3},
where $R_\text{X-Z}$ is the resistance associated with the loss/gain of the spin system.   
The latter can be used to define the dimensionless ``magnetic loss'' $\bar{\eta} \equiv - Q_0 / Q_\text{m} = R_\text{X-Z} / R_0$, which quantifies how strongly the TE$_{01\delta}$ mode interacts with the spin-polarized molecules compared to dissipative mode-cavity interactions associated with dielectric and conductive losses.
Here,  $\eta = -\bar{\eta}$ is the dimensionless maser gain (also known as the ``cooperativity'') considered in ref.~\cite{oxborrow2017maser};  $\eta > 1$ corresponds to the condition for maser oscillation for an unloaded cavity ($\eta > 2$ if critically loaded). As the last term in Eq.~\ref{eq:qdot} is associated with the (rate of change of) the number of photons in  the mode as a result of the mode's interaction with the spin system, the rate term $B(N_\textrm{X}-N_\textrm{Z})$ can be equated with $\omega_\textrm{mode}/Q_\textrm{m}$. The magnetic loss can thus be evaluated through $\bar{\eta} = Q_0 B (N_\textrm{Z}-N_\textrm{X}) / \omega_\textrm{mode}$.
Optical pumping thus leads to impedance mismatch, and hence reflection off the cavity coupling, quantified by
\begin{equation}
    \Gamma_\textrm{c} = -\frac{\bar{\eta}}{2 + \bar{\eta}};
\label{eq:eqS15}
\end{equation}
this relation is obtained (by eliminating $k$) from the following two relationships: $\Gamma_\textrm{c}=(k-1)/(k+1)$ and $k = (Q_0^{-1}-Q_\textrm{m}^{-1})^{-1}/Q_\textrm{ex}$,
where $k$ is the coupling coefficient of the cavity and $Q_0=Q_\textrm{ex}$ in our (ambiently critically coupled) set-up. 

As shown in Fig.~4(b) of the main text, the nadirs of  masar cooling correspond to turning points (minima) of the time-dependent photon number $q(t)$, i.e. $\dot{q}=0$. Based on Eq.~\ref{eq:qdot} and defining $T_\textrm{mode} = q/\epsilon = qhf_\textrm{mode}/k_\textrm{B}$, we obtain a relationship between $T_\textrm{mode}$ and $\bar{\eta}$:
\begin{equation}\label{eq:S16}
    T_\textrm{mode}=\frac{2T_0+\bar{\eta}T_\textrm{X-Z}}{2+\bar{\eta}}.
\end{equation}
Because during masar cooling, $T_\textrm{X-Z}$ is always significantly smaller than $T_0$, Eq.~\ref{eq:S16} can be further simplified to
\begin{equation}\label{eq:S17}
    T_\textrm{mode}=\frac{2T_0}{2+\bar{\eta}}.
\end{equation}
Combining Eq.~\ref{eq:eqS15} and Eq.~\ref{eq:S17}, the reflection coefficient of the cavity $\Gamma_\textrm{c}$ can thus be expressed in terms of the noise temperature of the microwave mode $T_\textrm{mode}$ through:
\begin{equation}\label{eq:eqS18}
    \Gamma_\textrm{c}=\frac{T_\textrm{mode}}{T_0}-1. 
\end{equation}
Therefore, by substituting Eq.~\ref{eq:eqS18} into Eq.~\ref{eq:S14}, the power reduction at the receiver's output resulting from masar cooling ($\Delta P$) can be expressed solely as a function of one (time-dependent) variable, namely $T_\textrm{mode}$. The resultant relation is plotted out in Fig.~4(c) of the main text.

\begin{figure}[htbp!]
\includegraphics[scale=0.5]{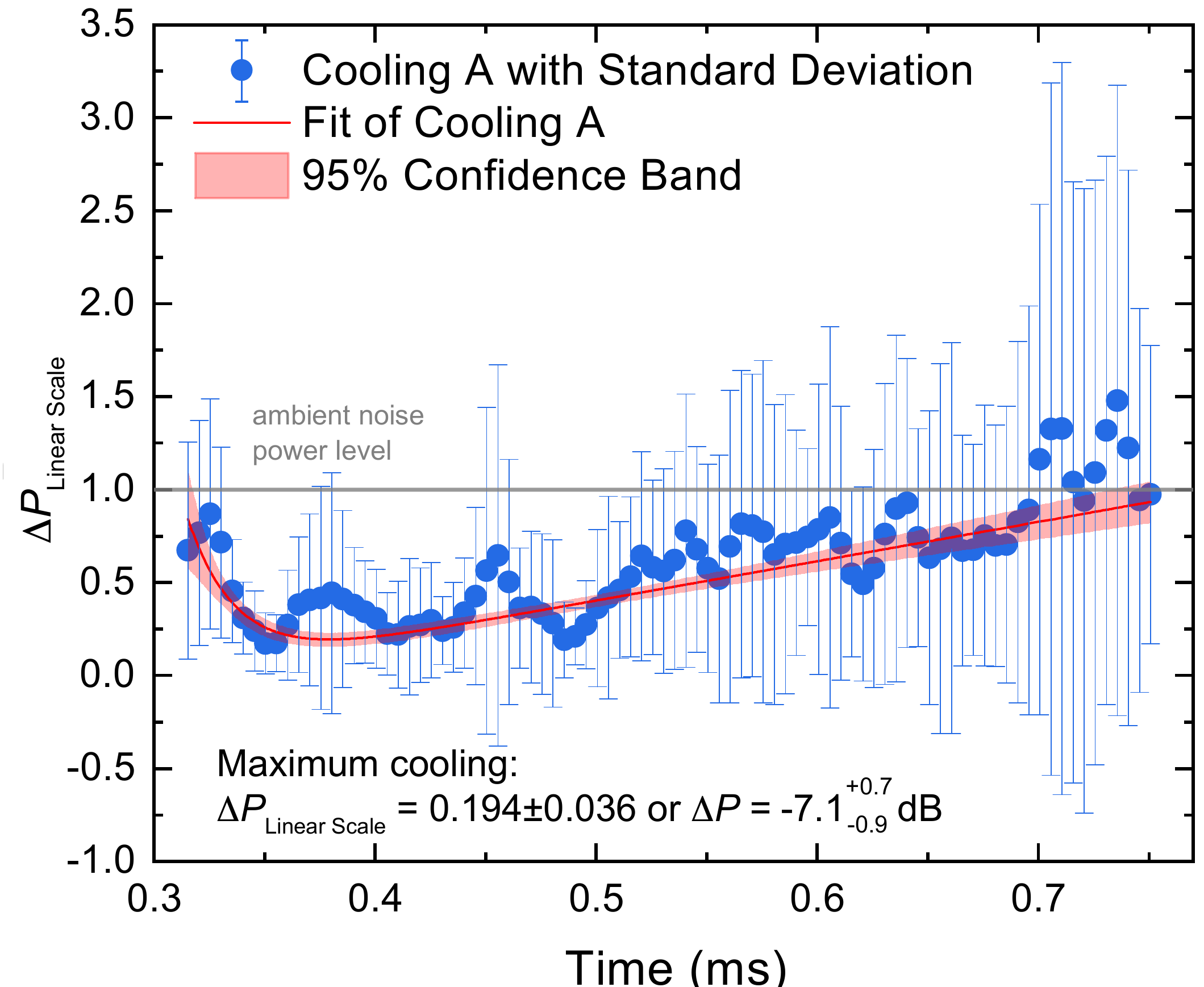}% Here is how to import EPS art
\caption{\label{fig:S4}Error analysis and fit of the Cooling A signal shown in Fig.~4(a) in the main text for determining the maximum cooling depth. The Cooling A signal with its standard deviation was obtained via average of 11 consecutive measurements. The blue dots represent the average of the Cooling A signal.}
\end{figure}

The maximum cooling, indicated by the lowest level of the Cooling A signal in Fig.~4(a), was obtained by fitting the signal data with a bi-exponential function. The best-fit curve with a 95\% confidence band is shown in Fig.~\ref{fig:S4}.

%\nocite{*}

%\bibliography{SMbib}% Produces the bibliography via BibTeX.
\providecommand{\noopsort}[1]{}\providecommand{\singleletter}[1]{#1}%